\documentclass[12pt]{article}

\usepackage{cite}
\usepackage{graphicx}
\usepackage{amssymb}
\usepackage{amsfonts}
\usepackage{amsmath}
\usepackage{longtable}
\usepackage{rotating}
\usepackage{lscape}
\usepackage{multirow}

\usepackage{hyperref}

\usepackage{fullpage}

\begin{document}

\def\ahad{a_\mu^{\mathrm{had, LO}}}
\def\asm{a^{\mathrm{SM}}_\mu}
\def\aexp{a^{\mathrm{exp}}_\mu}

\title{A comment on the impact of CMD-3 $e^+e^- \to \pi^+\pi^-$ cross section measurement on the SM $g_\mu-2$ value}

\author{V.\ V.\ Bryzgalov\footnote{Valery.Bryzgalov@ihep.ru}\, and\, 
        O.\ V.\ Zenin\footnote{zenin\_o@ihep.ru}\\
		{\it NRC ``Kurchatov Institute'' -- IHEP, Protvino, Russia}}
\date{}
\maketitle

\begin{abstract}
We estimated an impact of the recent CMD-3 measurement of the $e^+e^- \to \pi^+\pi^-$ total cross section at $0.3 < \sqrt{s} < 1.2$~GeV
on the leading order hadronic contribution $\ahad$ to the muon anomalous magnetic moment $a_\mu = (g_\mu-2)/2$, 
in presence of comparably precise ISR measurements of the cross section by BaBar and KLOE experiments being in significant tension with the CMD-3.
Assuming that all the experiments are affected by yet unidentified systematic effects, to account for the latter,   
we scaled the experimental uncertainties following the PDG prescription, 
thus facilitating a consistent joint fit of the world data on the $e^+e^- \to \pi^+\pi^-$ total cross section.
The same procedure was applied in all $e^+e^- \to hadrons$ channels contributing to the dispersive estimate of $\ahad$.
Despite an inclusion of the new CMD-3 $\pi^+\pi^-$ data, our estimate
$\ahad(e^+e^-) = (696.2 \pm 2.9) \times 10^{-10}$ is consistent with $\ahad(e^+e^-)$ values obtained by other authors before publication of the CMD-3 result. 
Including our $\ahad$ value into the SM prediction for $a_\mu$, 
we obtain $\asm = (11~659~184 \pm 4_{\mathrm{tot}}) \times 10^{-10}$ which is by $4.7\sigma$ smaller than the world average for the experimental value
$\aexp = (11~659~205.9 \pm 2.2) \times 10^{-10}$.
We confirm the observation by the CMD-3 authors that their $\sigma(e^+e^- \to \pi^+\pi^-)$ measurement, {\it when taken alone}, 
implies the $\asm$ prediction consistent with the $\aexp$ at $\sim 1\sigma$ level.
\end{abstract}

The $e^+e^- \to \pi^+\pi^-$ total cross section contributes $\simeq 70\%$ to the dispersive estimate~\cite{Petermann:1957ir} (see \cite{Aoyama:2020ynm} for a recent review) 
of the leading order hadronic term $\ahad$ in the SM prediction of the muon anomalous magnetic moment $\asm = (g_\mu - 2)/2$
and, on the other hand, introduces a major uncertainty in the $\asm$ determination.

\begin{figure}[hbp] 
  \begin{center}
  	\includegraphics[width=0.9\textwidth]{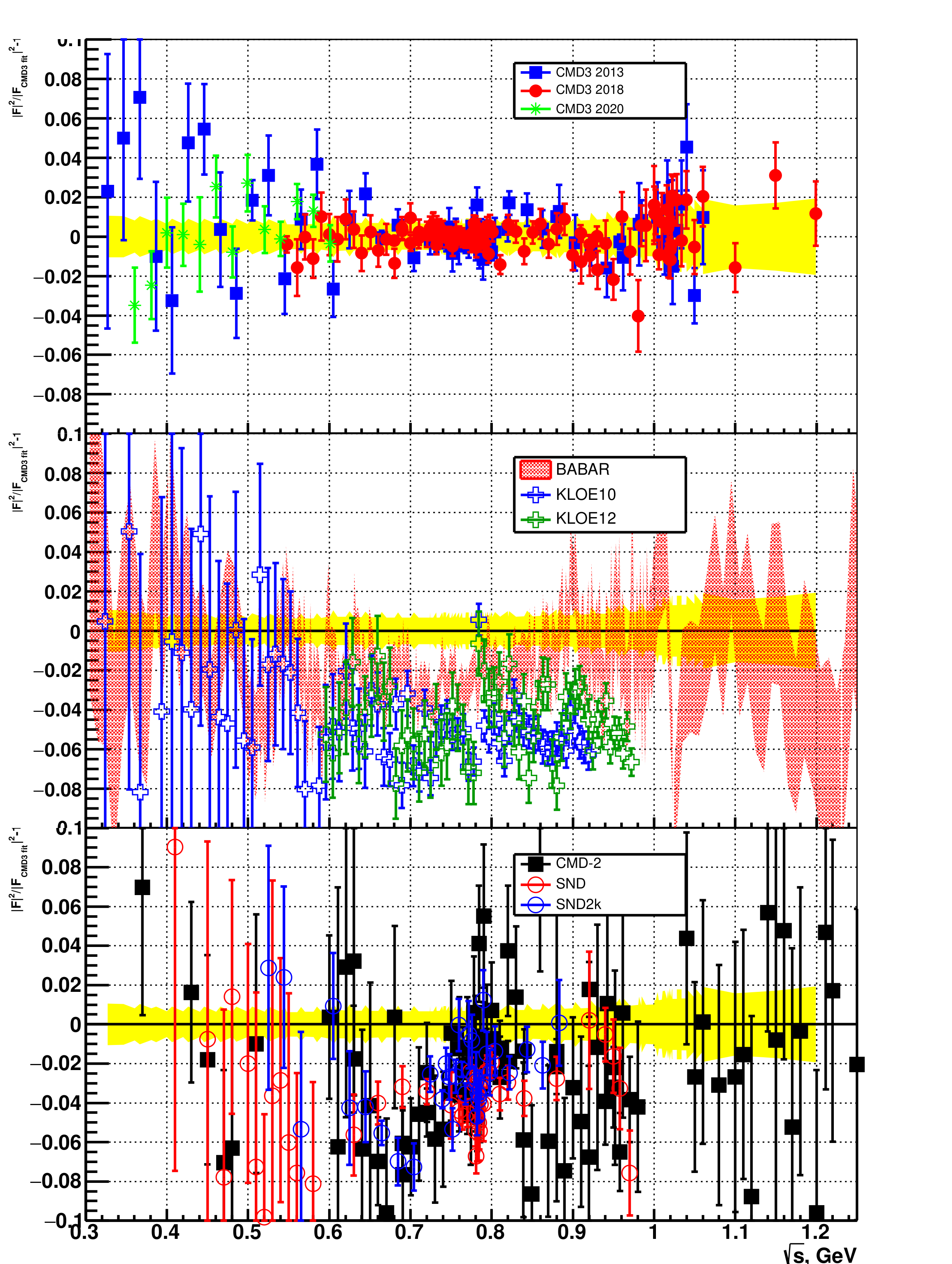}
  \end{center}
  \caption{\label{fig:cmd3-vs-others}
	\small
    (reprinted from \cite{CMD-3:2023rfe}) Relative differences between previous 
      measurements of the pion formfactor by
  	BaBar~\cite{BaBar:2012bdw}, 
  	KLOE~\cite{KLOE:2010qei,KLOE:2012anl} (middle plot),
  	CMD-2~\cite{CMD-2:2003gqi,CMD-2:2005mvb,Aulchenko:2006dxz,CMD-2:2006gxt}, 
  	SND~\cite{Achasov:2006vp} and SND2k~\cite{SND:2020nwa} (lower plot) experiments
  	and the fit (yellow band) to the CMD-3 data~\cite{CMD-3:2023alj,CMD-3:2023rfe} (upper plot).
  }
\end{figure}

Following the recent high precision measurement of the $e^+e^- \to \pi^+\pi^-$ cross section at $0.3 < \sqrt{s} < 1.2$~GeV by the CMD-3 collaboration~\cite{CMD-3:2023alj,CMD-3:2023rfe}
being in a significant tension with results previously published by 
BaBar~\cite{BaBar:2012bdw}, 
KLOE~\cite{KLOE:2010qei,KLOE:2012anl,KLOE-2:2017fda}, 
CMD-2~\cite{CMD-2:2003gqi,CMD-2:2005mvb,Aulchenko:2006dxz,CMD-2:2006gxt}, 
SND~\cite{Achasov:2006vp} and SND2k~\cite{SND:2020nwa} experiments (Fig.\,\ref{fig:cmd3-vs-others}),
we attempted to assess the impact of the CMD-3 result on the joint fit of the available $e^+e^- \to \pi^+\pi^-$ total cross section data and its implications for the $\asm$.
Details of the analysis were reported in~\cite{Bryzgalov:2023}.\footnote{The program code~\cite{code:2023} was used by the authors for the PDG mini-review 
``$\sigma$ and $R$ in $e^+e^-$ Collisions''~\cite{ParticleDataGroup:2022pth}. 
Essentially the same code was used for earlier revisions of the mini-review since~\cite{ParticleDataGroup:2002ivw}.
Details on the earliest version of the analysis are given in~\cite{Ezhela:2003pp,Kang:2002bm}
}

\begin{figure}[htbp] 
	\hspace*{-7ex}\includegraphics[width=1.25\textwidth]{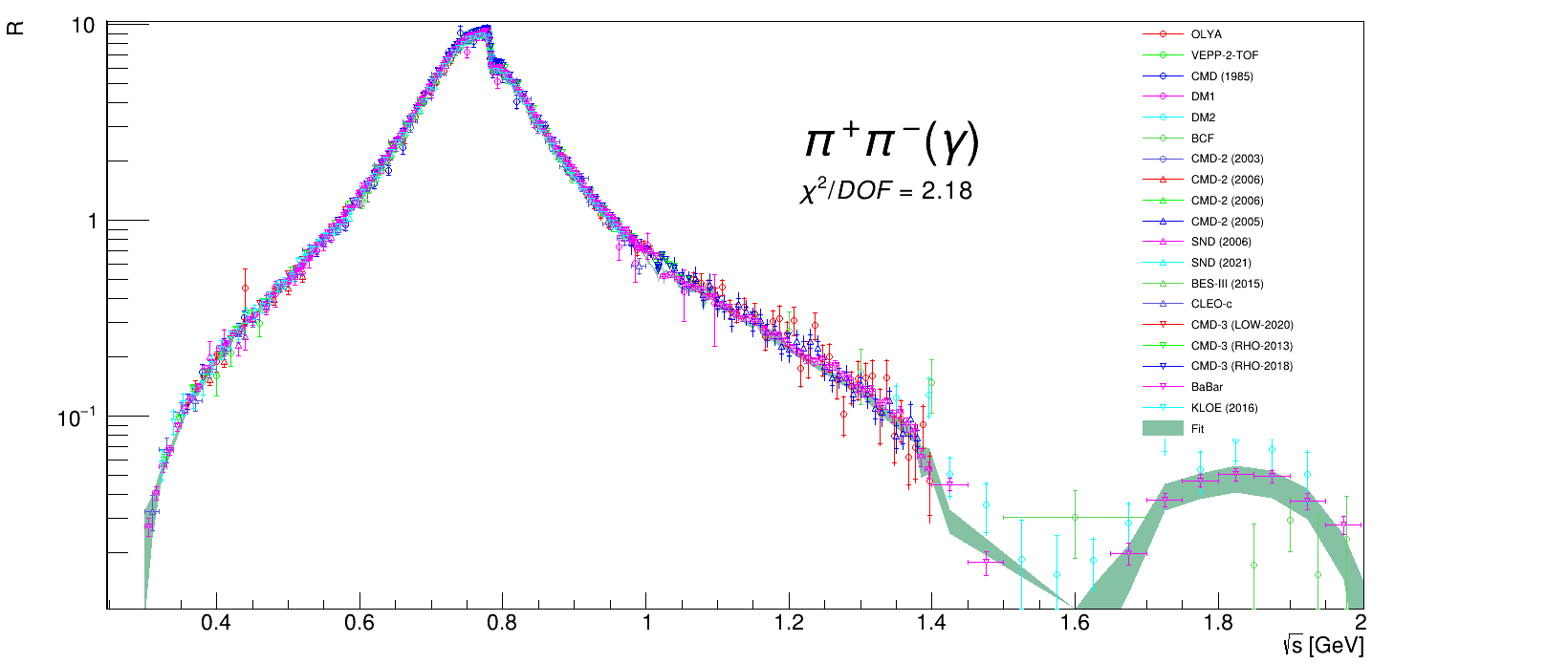}\\
		\hspace*{-6ex}\includegraphics[width=1.08\textwidth]{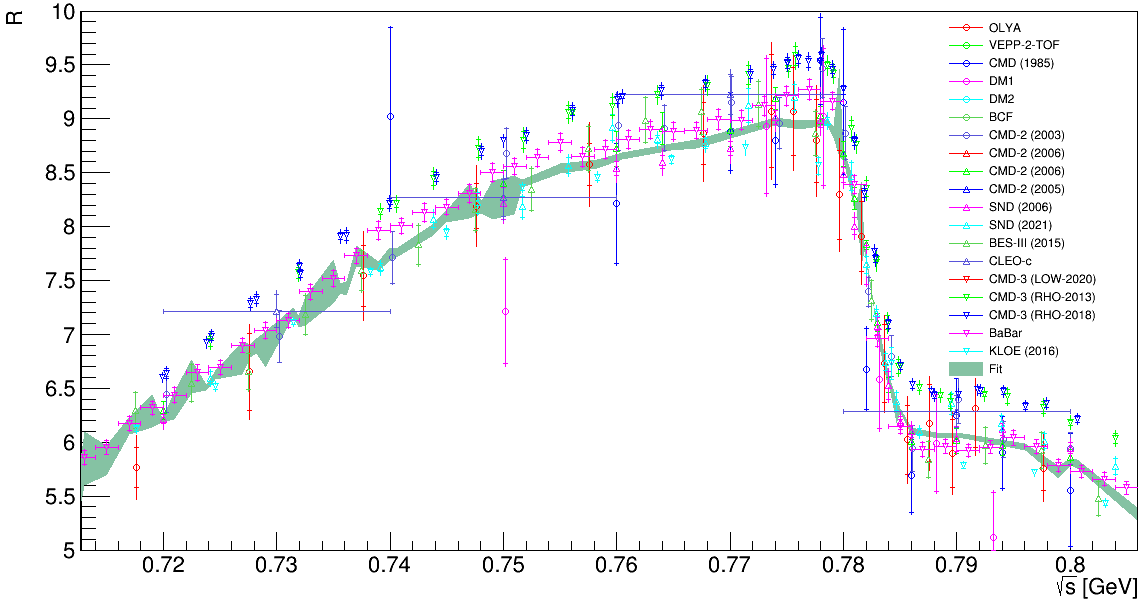}\\[-5ex]
	\caption{\label{fig:pipi} 
		\small
		The hadronic $R$-ratio\,\,$=\sigma(e^+e^- \to \pi^+\pi^-(\gamma), \mathrm{Born}) / \left(\frac{4\pi\alpha_0^2}{3s}\right)$.
		Statistical uncertainties of the data are shown by horizontal ticks on the vertical error bars.
		The latter correspond to the combined statistical and systematic uncertainties.
		Fit to the data by a continuous piecewise linear function is shown by the green band. See~\cite{Bryzgalov:2023} for details of the parameterization.
		The fit uncertainty is scaled by the $\sqrt{\chi^2/n_{\mathrm{dof}}}$ factor to account for incompatibility between individual experiments
		as explained in the text.
		The lower plot is an enlarged view of the $\rho$--$\omega$ interference region where incompatibility between CMD-3, BaBar and KLOE measurements 
		is most dramatic (cf. Fig.~\ref{fig:cmd3-vs-others}).
		References to the experimental data are listed in the $\pi^+\pi^-(\gamma)$ row of Table~\ref{table:a-mu-had-by-channel}.
	}
\end{figure}

\begin{figure}[htbp]
	\hspace*{-8ex}\includegraphics[width=1.1\textwidth]{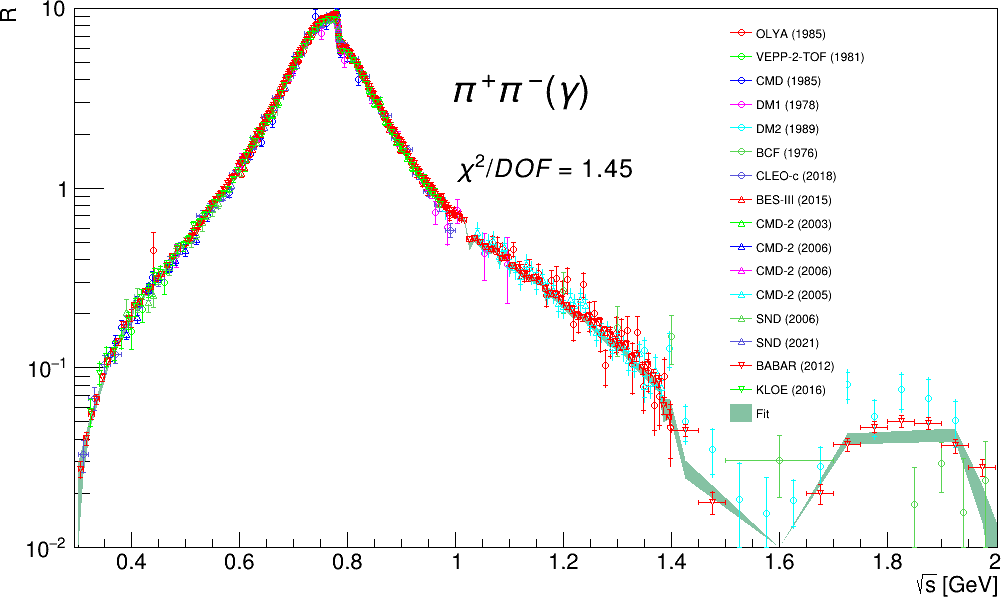}\\ 
	\hspace*{-7ex}\includegraphics[width=1.09\textwidth]{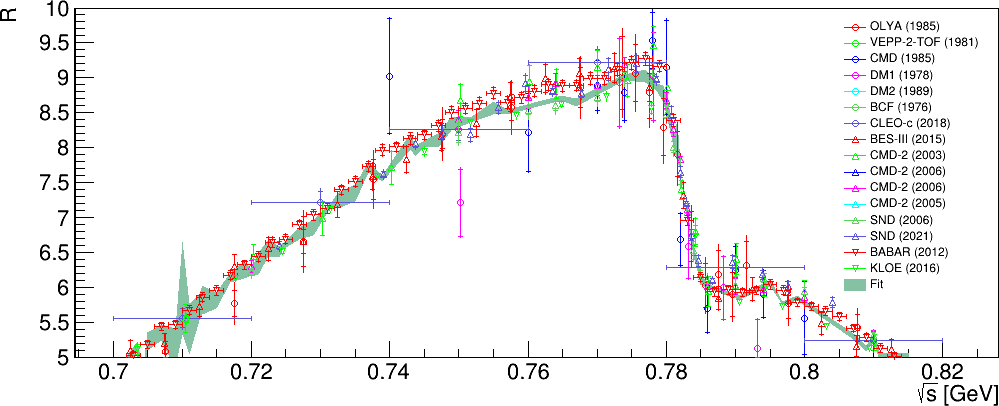}  
	\caption{\label{fig:no-cmd3}
			\small
			Same as Fig.\,\ref{fig:pipi} but without the CMD-3 $\pi^+\pi^-$ data.
			The lower plot demonstrates an inconsistency between BaBar and KLOE measurements.
			}
\end{figure}

Given that possible origins of a mutual systematic bias were not yet identified for the experiments measuring $\sigma(e^+e^- \to \pi^+\pi^-)$ at $\sim 1\%$ precision,
all of them must be treated {\it as is} on an equal basis.
From statistical viewpoint, this results in a low probability of the joint fit with $\chi^2/n_{\mathrm{dof}} = 2.18$ (Fig.\,\ref{fig:pipi}).
Before inclusion of the ``high'' CMD-3 data, the major source of statistical tension was the discrepancy between ISR based measurements of the cross section 
by BaBar~\cite{BaBar:2012bdw} and KLOE~\cite{KLOE-2:2017fda} experiments leading to the joint fit with $\chi^2/n_{\mathrm{dof}} = 1.45$ (Fig.\,\ref{fig:no-cmd3}).
For $n_{\mathrm{dof}} \gg 1$, a consistent fit should yield 
$\chi^2/n_{\mathrm{dof}} \in (1 - 2\sqrt{2/n_{\mathrm{dof}}}, 1 + 2\sqrt{2/n_{\mathrm{dof}}})$ 
with $\simeq 95\%$ probability. 
$\chi^2/n_{\mathrm{dof}}$ values outside this range indicate either an inadequate parameterization of the cross section 
or incompatibility between cross section values measured by different experiments.
As our model-independent parameterization of the cross section is constructed so that fitting it to mutually compatible measurements
should always give $\chi^2/n_{\mathrm{dof}} \simeq 1$~\cite{Bryzgalov:2023},
the poor fit is mostly due to incompatibility between BaBar, KLOE and CMD-3 experiments 
discussed in~\cite{CMD-3:2023rfe}.\footnote{This situation occurs not only in the $\pi^+\pi^-$ channel, see $\chi^2/{\mathrm{dof}}$ column in the Table~\ref{table:a-mu-had-by-channel}.}

A natural approach to treatment of incompatible measurements is to assume that each experiment 
is affected by randomly distributed systematic bias that cannot be identified within an isolated experiment and thus unaccounted for by its systematic uncertainty.
However, missing systematics can be assessed by observing the distribution of measurements of the same physical quantity made by several independent experiments.
The missing systematic uncertainty affecting all the experiments becomes an extra free parameter in the joint fit, with the optimum value maximizing the probability of the fit.
A widely used (though debated) PDG prescription~\cite{ParticleDataGroup:2022pth} is to apply a common scale factor\,\footnote{Dubbed the Birge factor in literature.} 
to error matrices of all experiments so that the fit yields $\chi^2/n_{\mathrm{dof}} = 1$. 
Being aware of shortcomings of this method, we still use it in our fits for the moment.
Effectively, the error matrix of each experiment must be scaled by $(\chi^2/n_{\mathrm{dof}})^{orig.}$ obtained in the fit with the unmodified error matrices.
At that, the central value of the fitted average cross section remains unchanged, while its uncertainty is 
scaled by the factor of $\sqrt{(\chi^2/n_{\mathrm{dof}})^{orig}}$.

The result of the joint fit to the $\sigma(e^+e^- \to \pi^+\pi^-)$ data at $0.3 < \sqrt{s} < 2$~GeV including CMD-3 is shown by the green band in Fig.\,\ref{fig:pipi}. 
The fit with unmodified experimental uncertainties gives $\chi^2/n_{\mathrm{dof}} = 2.18$ due to dramatic tension between BaBar, KLOE and CMD-3 measurements
at $0.6 < \sqrt{s} < 1.0$~GeV.\footnote{The fit restricted to the $0.6 < \sqrt{s} < 1.0$~GeV range would have much larger $\chi^2/n_{\mathrm{dof}}$ (see the lower plot in Fig.\,\ref{fig:pipi}).}
As explained above, all experimental uncertainties were inflated by $\sqrt{2.18}$ factor to make these measurements compatible.
On the plots, the individual measurements are shown with their original uncertainties while the fit result is shown with the uncertainty scaled by $\sqrt{2.18}$.
Scaling up the fit uncertainty accounts for incompatibility between the experiments, as well as for suboptimal parameterization of the cross section.

Substituting the fitted cross section into the dispersion integral for $\ahad$~\cite{Petermann:1957ir} we obtain the contribution of the $\pi^+\pi^-$ channel in the 
$0.3 < \sqrt{s} < 1.937$~GeV range:
\begin{equation}\label{eq:ahadpipi}
	\ahad(e^+e^- \to \pi^+\pi^-, 0.3 \div 1.937~\mathrm{GeV}) = (505.1 \pm 1.4_{\mathrm{exp.} e^+e^-} \pm 1.6_{\mathrm{par.}} \pm 0.6_{\mathrm{rad.}}) \times 10^{-10}\, ,
\end{equation}
where the first uncertainty is due to experimental uncertainties of the $\sigma(e^+e^- \to \pi^+\pi^-)$ data,
the second is the systematic uncertainty due to our cross section parameterization, 
and the last one is the uncertainty due to radiative corrections applied to the $e^+e^- \to \pi^+\pi^-$ data.

Results of application of the same procedure to other hadronic final states are shown in the Table~\ref{table:a-mu-had-by-channel}.
The total leading order hadronic contribution to $\asm$ is then 
\begin{equation}\label{eq:ahad}
	\ahad(e^+e^-) = (696.2 \pm 1.9_{\mathrm{exp.} e^+e^-} \pm 2.0_{\mathrm{par.}} \pm 0.8_{\mathrm{rad.}}) \times 10^{-10}\,.
\end{equation}
Despite an inclusion of the ``high'' CMD-3 $\sigma(e^+e^- \to \pi^+\pi^-)$ measurement, 
our estimate is still consistent with results obtained by the dispersive method by other authors using only pre-2021 $e^+e^-$ data~\cite{Aoyama:2020ynm}.
Ref.~\cite{Aoyama:2020ynm} quotes an average value 
$\ahad(e^+e^-) = (693.1 \pm 4.0_{\mathrm{tot}}) \times 10^{-10}$ obtained by merging results from 
Refs.~\cite{Davier:2019can,Davier:2017zfy,Keshavarzi:2019abf,Keshavarzi:2018mgv,Hoferichter:2019mqg,Colangelo:2018mtw}.\footnote{We also have a good per final state agreement with \cite{Keshavarzi:2019abf}.}
The SM prediction for $a_\mu$ including QED, electroweak, hadronic light-by-light, and NLO hadronic vacuum polarization contributions~\cite{Aoyama:2020ynm} and our $\ahad$ value (Eq.~\ref{eq:ahad}) is
\begin{equation}\label{eq:amu}
	\asm = (11~659~184 \pm 4_{\mathrm{tot}}) \times 10^{-10} \,,
\end{equation}
which is smaller than the experimental value $\aexp = (11~659~205.9 \pm 2.2) \times 10^{-10}$~\cite{Muong-2:2023cdq} by $4.7\sigma$.

An exclusion of the CMD-3 $\pi^+\pi^-$ data (Fig.\,\ref{fig:no-cmd3}) leads to a lower
$\ahad(e^+e^-) = (694.0 \pm 2.0_{\mathrm{exp.} e^+e^-} \pm 1.4_{\mathrm{par.}} \pm 0.4_{\mathrm{rad}}) \times 10^{-10}$ 
which translates to $\asm = (11659182 \pm 4_{\mathrm{tot}}) \times 10^{-10}$, lower than  $\aexp$ by $5\sigma$.

\begin{figure}[htbp]
	\hspace*{-8ex}\includegraphics[width=1.1\textwidth]{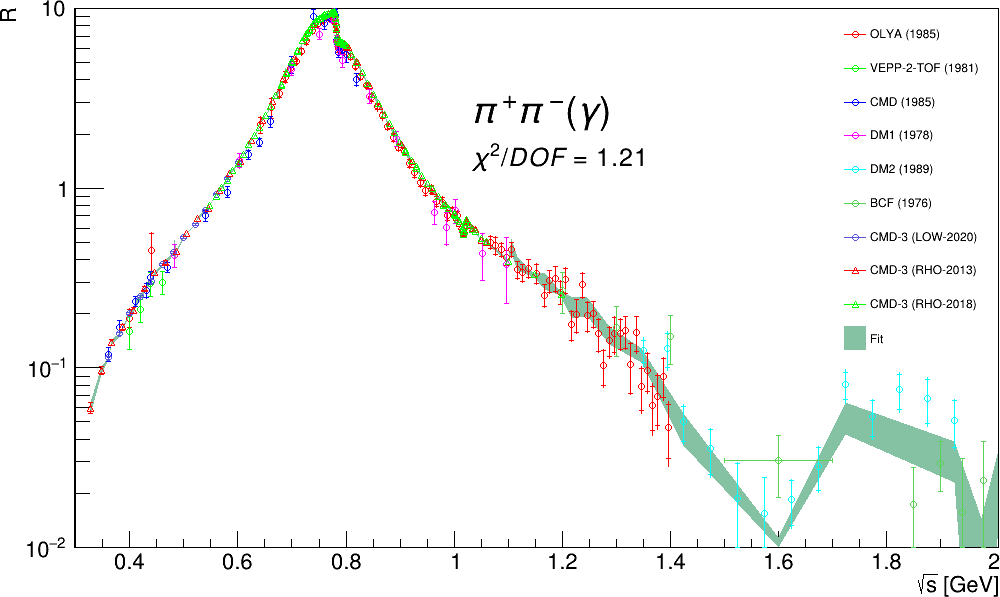}\\ 
	\hspace*{-7ex}\includegraphics[width=1.09\textwidth]{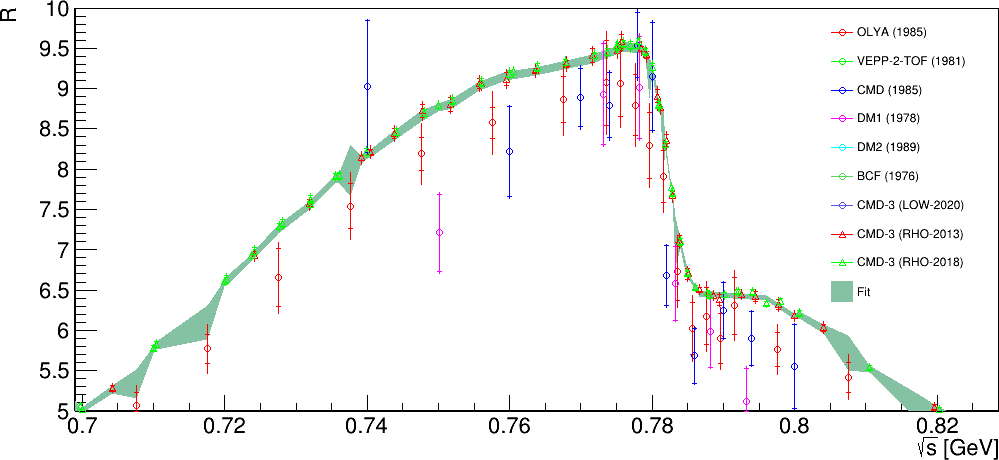}\\[-3ex] 
	\caption{\label{fig:only-cmd3}
		\small
		Same as Fig.\,\ref{fig:pipi} but without experiments measuring $\sigma(e^+e^- \to \pi^+\pi^-)$ to $\sim 1\%$ precision except CMD-3.
		This makes the fit effectively dominated by CMD-3 at $\sqrt{s} < 1.2$~GeV.
		The fit results in  
		$\ahad(e^+e^- \to \pi^+\pi^-, 0.3 \div 1.937~\mathrm{GeV}) = (529.6 \pm 2.8_{\mathrm{exp.} e^+e^-} \pm 3.3_{\mathrm{par.}} \pm 3.3_{\mathrm{rad.}}) \times 10^{-10}$ 
		corresponding to the total $\ahad = (723.4 \pm 3.1_{\mathrm{exp.}} \pm 3.1_{\mathrm{par.}} \pm 3.5_{\mathrm{rad.}})\times 10^{-10}$
		and $\asm = (11~659~211 \pm 6_{\mathrm{tot}}) \times 10^{-10}$ consistent with the experimental value $\aexp = (11~659~205.9 \pm 2.2) \times 10^{-10}$, 
		as first pointed out in the CMD-3 paper~\cite{CMD-3:2023rfe}.
	}
\end{figure}

An extreme exercise would be to estimate an impact of the CMD-3 $\sigma(e^+e^- \to \pi^+\pi^-)$ measurement taken as the dominant experimental input to $\ahad(\pi^+\pi^-)$ 
at $0.3 < \sqrt{s} < 1.2$~GeV. 
For this purpose, we completely excluded from the fit all the experiments measuring the cross section with $\sim 1\%$ precision, except CMD-3
(Fig.\,\ref{fig:only-cmd3}).\footnote{In a ``CMD-3 only'' fit, retaining BaBar data in the $\sqrt{s} > 1.2$~GeV range uncovered by CMD-3 would be methodically incorrect due to strong correlations 
between BaBar data points at $\sqrt{s} < 1.2$~GeV (where we ignore BaBar) and $\sqrt{s} > 1.2$~GeV.
Note that in this case the cross section parameterization uncertainty for the total $\ahad$ is slightly lower than the one in the $\pi^+\pi^-$ channel due to anti-correlation between systematic variations in different channels.}
This gives 
$\ahad(e^+e^- \to \pi^+\pi^-, 0.3 \div 1.937~\mathrm{GeV}) = (529.6 \pm 2.8_{\mathrm{exp.} e^+e^-} \pm 3.3_{\mathrm{par.}} \pm 3.3_{\mathrm{rad.}}) \times 10^{-10}$ 
which corresponds to the total
$\ahad = (723.4 \pm 3.1_{\mathrm{exp.}} \pm 3.1_{\mathrm{par.}} \pm 3.5_{\mathrm{rad.}})\times 10^{-10}$
and, hence, to $\asm = (11~659~211 \pm 6_{\mathrm{tot}}) \times 10^{-10}$ 
consistent with the experimental value $\aexp = (11~659~205.9 \pm 2.2) \times 10^{-10}$, as pointed out in Ref.~\cite{CMD-3:2023rfe}.

\begin{figure}[htbp] 
	\hspace*{-10ex}\includegraphics[width=1.23\textwidth]{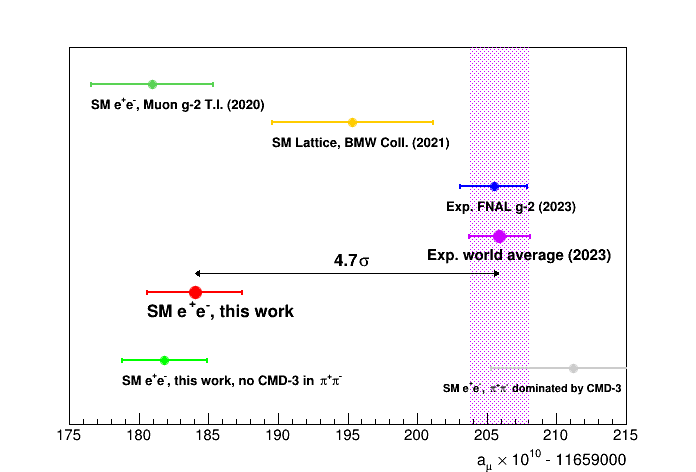}
	\caption{\label{fig:asm-vs-exp}%
	\small
	Summary of $\asm$ estimates vs the $\aexp$ world average currently dominated by the FNAL $g_\mu-2$ measurement~\cite{Muong-2:2023cdq}.
	The plot shows the $\asm$ values including our nominal (red) $\ahad$ estimate utilizing $e^+e^- \to hadrons$ data including CMD-3,
	our $\ahad$ estimate without CMD-3 (green), and the tentative $\ahad$ value with the $\pi^+\pi^-$ channel dominated by CMD-3 (grey).
	The $\asm$ value including the dispersive $\ahad$ value by the Muon $g-2$ Theory Initiative~\cite{Aoyama:2020ynm} is shown in dark green.
	The $\asm$ including lattice QCD calculation of $\ahad$  by the BMW Collaboration~\cite{Borsanyi:2020mff} is shown in orange.
	}
\end{figure}

Summary of $\asm$ estimates including the above results confronted with the experimental $\aexp$ value is shown in Fig.\,\ref{fig:asm-vs-exp}.

\vspace*{0.5ex}

{\bf In conclusion}, 
we estimated an impact of the recent CMD-3 measurement of the $e^+e^- \to \pi^+\pi^-$ total cross section at $0.3 < \sqrt{s} < 1.2$~GeV
on the leading order hadronic contribution $\ahad$ to the muon anomalous magnetic moment $a_\mu = (g_\mu-2)/2$, 
in presence of comparably precise ISR measurements of the cross section by BaBar and KLOE experiments being in significant tension with the CMD-3.
Assuming that all the experiments are affected by yet unidentified systematic effects, to account for the latter,   
we scaled the experimental uncertainties following the PDG prescription, 
thus facilitating a consistent joint fit of the world data on the $e^+e^- \to \pi^+\pi^-$ total cross section.
The same procedure was applied in all $e^+e^- \to hadrons$ channels contributing to the dispersive estimate of $\ahad$.
Despite an inclusion of the new CMD-3 data, our estimate
$\ahad(e^+e^-) = (696.2 \pm 2.9) \times 10^{-10}$ is consistent with $\ahad(e^+e^-)$ values obtained by other authors before publication of the CMD-3 result. 
Including our $\ahad$ value into the SM prediction for $a_\mu$, 
we obtain $\asm = (11~659~184 \pm 4_{\mathrm{tot}}) \times 10^{-10}$ which is by $4.7\sigma$ smaller than the world average for the experimental value
$\aexp = (11~659~205.9 \pm 2.2) \times 10^{-10}$.
We confirm the observation made in the CMD-3 paper~\cite{CMD-3:2023rfe} that CMD-3 $\pi^+\pi^-$ data, when taken alone, 
imply the $\asm$ prediction consistent with the $\aexp$ at $\sim 1\sigma$ level.

\paragraph*{Acknowledgements.}
The authors are indebted to V.~B.~Anikeev and \framebox{S.~I.~Bityukov} for valuable discussions on statistical matters.

\begin{table}[htbp]
	\scriptsize
	\begingroup
\renewcommand{\arraystretch}{1.2}
\begin{tabular}{lrrrl}
		Final state & \parbox{0.24\textwidth}{$a_\mu(\mathrm{had},\mathrm{LO})$ $\times 10^{10}$\\ { \phantom{000.000} (exp.) (par.) (rad.)}}  & $\sqrt{s}\, [\mathrm{GeV}]$ & $\chi^2/\mathrm{dof}$ & References \\[0.5ex]
\hline
\hline
$ \pi^+ \pi^- (\gamma) $ & 505.147 (1.367) (1.551) (0.606) & 0.3 $ \div $ 1.937 & 2.18 & \parbox{28ex}{\cite{BESIII:2015equ,CMD-3:2023alj,CMD-2:2003gqi,Achasov:2006vp,Aulchenko:2006dxz,CMD-2:2006gxt,KLOE-2:2017fda,SND:2020nwa,Bollini:1975nw,Barkov:1985ac,DM2:1988xqd,Quenzer:1978qt,BaBar:2012bdw,Xiao:2017dqv,Vasserman:1981xq,CMD-2:2005mvb}} \\
$ \pi^+ \pi^- \pi^0 $ & 48.481 (0.967) (0.629) (0.066) & 0.66 $ \div $ 1.937 & 1.79 & \parbox{28ex}{\cite{Achasov:2002ud,CMD-2:2003gqi,CMD:1989aaa,Cordier:1979qg,Akhmetshin:1995vz,Akhmetshin:1998se,Achasov:2000am,Achasov:2003ir,BaBar:2004ytv,Dolinsky:1991vq,Aulchenko:2015mwt,DM2:1992zkc}} \\
$ \pi^+ \pi^- 2\pi^0 $ & 18.778 (0.431) (0.509) (0.067) & 0.85 $ \div $ 1.937 & 1.94 & \parbox{28ex}{\cite{SND:2001mhi,DM2:1990kxb,Esposito:1981dv,Cosme:1978qe,Bacci:1980zs,CMD-2:1998gab,Cosme:1976tf,BaBar:2017zmc,Dolinsky:1991vq,Kurdadze:1986tc}} \\
$ 2\pi^+ 2\pi^- $ & 15.397 (0.181) (0.060) (0.043) & 0.6125 $ \div $ 1.937 & 2.34 & \parbox{28ex}{\cite{SND:2001mhi,DM2:1990kxb,Cosme:1978qe,Cordier:1981zn,CMD-2:1999dof,CMD-2:2000rfr,CMD-2:2004vyr,Cosme:1976tf,Akhmetshin:2016dtr,Cordier:1978yp,Bacci:1980ru,BaBar:2012sxt,Dolinsky:1991vq,Barkov:1988gp,Kurdadze:1988mu}} \\
$ K^+ K^- $ & 23.211 (0.188) (0.072) (0.009) & 0.985 $ \div $ 1.937 & 1.99 & \parbox{28ex}{\cite{CMD-2:2008fsu,Kozyrev:2017agm,CMD:1983aaa,Esposito:1980bz,Ivanov:1981wf,Delcourt:1980eq,Achasov:2000am,BaBar:2013jqz,Achasov:2016lbc}} \\
$ K_S K_L $ & 13.188 (0.130) (0.000) (0.000) & 1.00371 $ \div $ 1.937 & 0.95 & \parbox{28ex}{\cite{CMD-2:1999chh,CMD:1983aaa,Akhmetshin:1995vz,Akhmetshin:2002vj,CMD-3:2016nhy,Mane:1980ep,Achasov:2000am,BaBar:2014uwz,Ivanov:1982cr}} \\
$ \pi^0 \gamma $ & 4.359 (0.093) (0.049) (0.000) & 0.59986 $ \div $ 1.38 & 1.70 & \parbox{28ex}{\cite{Achasov:2000zd,CMD-2:2004ahv,SND:2016drm}} \\
$ K_S K^+ \pi^- + K_S K^- \pi^+ $ & 1.814 (0.100) (0.000) (0.000) & 1.24 $ \div $ 1.937 & 0.99 & \parbox{28ex}{\cite{BaBar:2007ceh}} \\
$ 2\pi^+ 2\pi^- \pi^0 $ & 1.746 (0.043) (0.000) (0.009) & 1.0125 $ \div $ 1.937 & 0.00 & \parbox{28ex}{\cite{Cosme:1978qe,BaBar:2007qju,Barkov:1988gp}} \\
$ 2\pi^+ 2\pi^0 2\pi^- $ & 1.728 (0.198) (0.034) (0.000) & 1.3125 $ \div $ 1.937 & 1.99 & \parbox{28ex}{\cite{BaBar:2006vzy,Schioppa:1986aaa,Barkov:1988gp}} \\
$ 2\pi^+ 2\pi^- 3\pi^0 $ & 0.099 (0.013) (0.002) (0.001) & 1.575 $ \div $ 1.937 & 0.57 & \parbox{28ex}{\cite{BaBar:2021rki}} \\
$ 3\pi^+ 3\pi^- $ & 0.240 (0.014) (0.000) (0.012) & 1.3125 $ \div $ 1.937 & 0.00 & \parbox{28ex}{\cite{BaBar:2006vzy,Bisello:1981sh,CMD-3:2013nph,Schioppa:1986aaa,Barkov:1988gp}} \\
$ 3\pi^+ 3\pi^- \pi^0 $ & 0.020 (0.004) (0.001) (0.000) & 1.6 $ \div $ 1.937 & 0.65 & \parbox{28ex}{\cite{CMD-3:2019ufp}} \\
$ \eta \gamma $ & 0.691 (0.051) (0.000) (0.000) & 0.6 $ \div $ 1.354 & 1.36 & \parbox{28ex}{\cite{Achasov:2000zd,Akhmetshin:1995vz,CMD-2:2001dnv,Cosme:1975rs}} \\
$ \eta \pi^+ \pi^- $ & 0.575 (0.019) (0.000) (0.000) & 1.15 $ \div $ 1.937 & 1.18 & \parbox{28ex}{\cite{CMD-2:2000mlo,BaBar:2007qju,SND:2014rfi,BaBar:2018erh,Dolinsky:1991vq}} \\
$ K^+ K^- \pi^0 $ & 0.202 (0.050) (0.000) (0.001) & 1.44 $ \div $ 1.937 & 0.54 & \parbox{28ex}{\cite{DM2:1990npw,Bisello:1991kd}} \\
$ K^+ K^- \pi^0 \pi^0 $ & 0.100 (0.011) (0.000) (0.000) & 1.5 $ \div $ 1.937 & 1.32 & \parbox{28ex}{\cite{BaBar:2011btv}} \\
$ K^+ K^- \pi^+ \pi^- $ & 0.799 (0.033) (0.000) (0.000) & 1.4 $ \div $ 1.937 & 0.00 & \parbox{28ex}{\cite{DM2:1990npw,Cordier:1981en,Shemyakin:2015cba,BaBar:2011btv}} \\
$ K^+ K^- \pi^+ \pi^- \pi^0 $ & 0.129 (0.024) (0.000) (0.000) & 1.6125 $ \div $ 1.937 & 1.63 & \parbox{28ex}{\cite{BaBar:2007qju}} \\
$ K_S K_L \eta $ & 0.238 (0.059) (0.000) (0.000) & 1.575 $ \div $ 1.937 & 1.31 & \parbox{28ex}{\cite{BaBar:2017nrz}} \\
$ K_S K_L \pi^0 $ & 0.839 (0.114) (0.000) (0.000) & 1.425 $ \div $ 1.937 & 1.50 & \parbox{28ex}{\cite{BaBar:2017nrz}} \\
$ K_S K_L \pi^0 \pi^0 $ & 0.137 (0.043) (0.000) (0.000) & 1.35 $ \div $ 1.937 & 0.00 & \parbox{28ex}{\cite{BaBar:2017nrz}} \\
$ K_S K_L \pi^+ \pi^- $ & 0.166 (0.028) (0.000) (0.000) & 1.425 $ \div $ 1.937 & 0.00 & \parbox{28ex}{\cite{BaBar:2014uwz}} \\
$ K_S K^+ \pi^- \pi^0 + K_S K^- \pi^+ \pi^0 $ & 0.640 (0.044) (0.000) (0.000) & 1.51 $ \div $ 1.937 & 1.08 & \parbox{28ex}{\cite{BaBar:2017pkz}} \\
$ K_S K_S \pi^+ \pi^- $ & 0.066 (0.007) (0.000) (0.000) & 1.63 $ \div $ 1.937 & 1.37 & \parbox{28ex}{\cite{BaBar:2014uwz}} \\
$ \omega(783) \eta $ & 0.035 (0.002) (0.000) (0.000) & 1.34 $ \div $ 1.937 & 0.85 & \parbox{28ex}{\cite{CMD-3:2017tgb,Achasov:2016qvd}} \\
$ \omega(783) < \pi^0 \gamma > \pi^0 $ & 0.894 (0.021) (0.000) (0.000) & 0.75 $ \div $ 1.937 & 1.56 & \parbox{28ex}{\cite{KLOE:2008woc,Aulchenko:2000mr,CMD-2:2003bgh,Achasov:2012zza,DM2:1990kxb,Achasov:1999wr,Dolinsky:1986kj,CMD-2:2003mfj,Achasov:2016zvn}} \\
$ \omega(783) < \pi^+ \pi^- \pi^0 > \pi^+ \pi^- $ & 0.098 (0.005) (0.000) (0.000) & 1.15 $ \div $ 1.937 & 1.10 & \parbox{28ex}{\cite{Cordier:1981zs,CMD-2:2000mlo,BaBar:2007qju,DM2:1992zkc}} \\
$ \omega \eta \pi^0 $ & 0.055 (0.043) (0.000) (0.000) & 1.5 $ \div $ 1.937 & 1.16 & \parbox{28ex}{\cite{Achasov:2016eyg}} \\
$ \phi(1020) \eta $ & 0.067 (0.003) (0.000) (0.000) & 1.56 $ \div $ 1.937 & 0.98 & \parbox{28ex}{\cite{Ivanov:2019crp,BaBar:2007qju,BaBar:2007ceh}} \\
$ \pi^+ \pi^- 2\pi^0 \eta $ & 0.117 (0.019) (0.000) (0.000) & 1.625 $ \div $ 1.937 & 0.85 & \parbox{28ex}{\cite{BaBar:2018rkc}} \\
$ \pi^+ \pi^- 3\pi^0 $ & 1.067 (0.112) (0.000) (0.000) & 1.125 $ \div $ 1.937 & 0.68 & \parbox{28ex}{\cite{BaBar:2018rkc}} \\
$ \pi^+ \pi^- \pi^0 \eta $ & 0.663 (0.075) (0.000) (0.000) & 1.394 $ \div $ 1.937 & 0.82 & \parbox{28ex}{\cite{CMD-3:2017tgb}} \\
$ p \bar p $ & 0.030 (0.001) (0.000) (0.000) & 1.889 $ \div $ 1.937 & 1.24 & \parbox{28ex}{\cite{Bisello:1983at,CMD-3:2015fvi,Delcourt:1979ed,BaBar:2013ves}} \\
$ n \bar n $ & 0.028 (0.006) (0.000) (0.000) & 1.89 $ \div $ 1.937 & 1.24 & \parbox{28ex}{\cite{Antonelli:1998fv,Achasov:2014ncd}} \\
$ 2hadron (hadrons) $ & 43.509 (0.722) (0.661) (0.000) & 1.937 $ \div $ 11.199 & 1.35 & \parbox{28ex}{\cite{BES:2009ejh,BES:2001ckj,Ablikim:2006mb,Barber:1981tc,Albrecht:1982bq,TASSO:1984dcd,CELLO:1986ijz,Venus:1987dew,TOPAZ:1989phk,VENUS:1990vwh,TOPAZ:1993iqj,TOPAZ:1994ymb,KEDR:2018hhr,Aachen-DESY-AnnecyLAPP-MIT-NIKHEF-Beijing:1979sxf,CLEO:1983ffg,Fernandez:1984yw,Mark-J:1986ave,AMY:1990xpt,MARK-II:1990awn,CLEO:1997eca,Rapidis:1977cv,Rice:1982br,BES:1999wbx,Naroska:1986si,Schindler:1979rj,Osterheld:1986hw,Edwards:1990pc,LENA:1982shw,TASSO:1983cre,CrystalBall:1988obw,TASSO:1990cdg,TASSO:1979rma,ARGUS:1991ztk,DESY-Hamburg-Heidelberg-Munich:1980ttv,Blinov:1993fw}} \\
\hline
pQCD & 2.065 \phantom{(0.000)} (0.002) \phantom{(0.000)}   & $ > $ 11.1990  &  & \\
ChPT $ \pi\pi, \pi^0\gamma $ & 0.538 \phantom{(0.000)} (0.013) \phantom{(0.000)}   &  0.2792  $ \div $ 0.3000  &  & \\
\hline
$ \Psi(1S) $ &  6.495 \phantom{(0.000)} (0.124)  \phantom{(0.000)} &  3.0969 & & \\
$ \Psi(2S) $ &  1.631 \phantom{(0.000)} (0.057)  \phantom{(0.000)} &  3.6861 & & \\
$ \Upsilon(1S) $ &  0.054 \phantom{(0.000)} (0.002)  \phantom{(0.000)} &  9.4604 & & \\
$ \Upsilon(2S) $ &  0.021 \phantom{(0.000)} (0.003)  \phantom{(0.000)} & 10.0234 & & \\
$ \Upsilon(3S) $ &  0.014 \phantom{(0.000)} (0.002)  \phantom{(0.000)} & 10.3551 & & \\
$ \Upsilon(4S) $ &  0.010 \phantom{(0.000)} (0.001)  \phantom{(0.000)} & 10.5794 & & \\
\hline
\hline
 Total & {\bf696.181} (1.925) (1.953) (0.813) &  &  & \\
\end{tabular}
\endgroup

	\vspace*{-1ex}
	\caption{\label{table:a-mu-had-by-channel} 
	\footnotesize
	Summary of $\ahad$ contributions from the exclusive $\sigma(e^+e^- \to hadrons)$ measurements at $\sqrt{s} < 1.937$~GeV and
	inclusive $\sigma(e^+e^- \to 2hadron (hadrons))$ measurements at $1.937 < \sqrt{s} < 11.199$~GeV. 
	The first uncertainty is the experimental one scaled 
	by $\sqrt{\chi^2/n_{\mathrm{dof}}}$ in channels with $\chi^2/n_{\mathrm{dof}} > 1$.
	The second uncertainty is due to our parameterization of the experimental cross section used in the $\ahad$ dispersion integral.
	The last uncertainty is due to radiative corrections applied to the experimental data.
	At $\sqrt{s} < 0.3$~GeV we use ChPT parameterization for $e^+e^- \to \pi^0\gamma$, $\pi^+\pi^-$ cross sections.
	The 3-loop pQCD parameterization of the total $e^+e^- \to hadrons$ cross section is used at $\sqrt{s} > 11.199$~GeV.
	Contributions of the narrow $\Psi(1,2S)$ and $\Upsilon(1-4S)$ resonances are accounted for using the Breit--Wigner parameterization.
	See details in Ref.~\cite{Bryzgalov:2023}.
	}
\end{table}
	
\newpage

\bibliographystyle{utphys}

\bibliography{refs}

\providecommand{\href}[2]{#2}\begingroup\raggedright\begin{thebibliography}{10%
0}

\bibitem{Petermann:1957ir}
A.~Petermann, ``{Magnetic moment of the $\mu$ meson},''
  \href{https://dx.doi.org/10.1103/PhysRev.105.1931}{{\em Phys. Rev.}
  {\bfseries 105} (1957) 1931}, CERN-57-27.

\bibitem{Aoyama:2020ynm}
T.~Aoyama {\em et~al.}, ``{The anomalous magnetic moment of the muon in the
  Standard Model},''
  \href{https://dx.doi.org/10.1016/j.physrep.2020.07.006}{{\em Phys. Rept.}
  {\bfseries 887} (2020) 1--166},
  \href{https://arxiv.org/abs/2006.04822}{{\ttfamily arXiv:2006.04822
  [hep-ph]}}.

\bibitem{CMD-3:2023alj}
{\bfseries CMD-3} Collaboration, F.~V. Ignatov {\em et~al.}, ``{Measurement of
  the $e^+e^-\to\pi^+\pi^-$ cross section from threshold to 1.2 GeV with the
  CMD-3 detector},'' \href{https://arxiv.org/abs/2302.08834}{{\ttfamily
  arXiv:2302.08834 [hep-ex]}}.

\bibitem{CMD-3:2023rfe}
{\bfseries CMD-3} Collaboration, F.~V. Ignatov {\em et~al.}, ``{Measurement of
  the pion formfactor with CMD-3 detector and its implication to the hadronic
  contribution to muon (g-2)},''
  \href{https://arxiv.org/abs/2309.12910}{{\ttfamily arXiv:2309.12910
  [hep-ex]}}.

\bibitem{BaBar:2012bdw}
{\bfseries BaBar} Collaboration, J.~P. Lees {\em et~al.}, ``{Precise
  Measurement of the $e^+ e^- \to \pi^+\pi^- (\gamma)$ Cross Section with the
  Initial-State Radiation Method at BABAR},''
  \href{https://dx.doi.org/10.1103/PhysRevD.86.032013}{{\em Phys. Rev. D}
  {\bfseries 86} (2012) 032013}, BABAR-PUB-12-003,
  \href{https://arxiv.org/abs/1205.2228}{{\ttfamily arXiv:1205.2228 [hep-ex]}}.

\bibitem{KLOE:2010qei}
{\bfseries KLOE} Collaboration, F.~Ambrosino {\em et~al.}, ``{Measurement of
  $\sigma(e^+ e^- \to \pi^+ \pi^-)$ from threshold to 0.85 GeV$^2$ using
  Initial State Radiation with the KLOE detector},''
  \href{https://dx.doi.org/10.1016/j.physletb.2011.04.055}{{\em Phys. Lett. B}
  {\bfseries 700} (2011) 102--110},
  \href{https://arxiv.org/abs/1006.5313}{{\ttfamily arXiv:1006.5313 [hep-ex]}}.

\bibitem{KLOE:2012anl}
{\bfseries KLOE} Collaboration, D.~Babusci {\em et~al.}, ``{Precision
  measurement of $\sigma(e^+e^-\rightarrow \pi^+\pi^-\gamma)/
  \sigma(e^+e^-\rightarrow \mu^+\mu^-\gamma)$ and determination of the
  $\pi^+\pi^-$ contribution to the muon anomaly with the KLOE detector},''
  \href{https://dx.doi.org/10.1016/j.physletb.2013.02.029}{{\em Phys. Lett. B}
  {\bfseries 720} (2013) 336--343},
  \href{https://arxiv.org/abs/1212.4524}{{\ttfamily arXiv:1212.4524 [hep-ex]}}.

\bibitem{KLOE-2:2017fda}
{\bfseries KLOE-2} Collaboration, A.~Anastasi {\em et~al.}, ``{Combination of
  KLOE $\sigma\big(e^+e^-\rightarrow\pi^+\pi^-\gamma(\gamma)\big)$ measurements
  and determination of $a_{\mu}^{\pi^+\pi^-}$ in the energy range $0.10 < s <
  0.95$ GeV$^2$},'' \href{https://dx.doi.org/10.1007/JHEP03(2018)173}{{\em
  JHEP} {\bfseries 03} (2018) 173},
  \href{https://arxiv.org/abs/1711.03085}{{\ttfamily arXiv:1711.03085
  [hep-ex]}}.

\bibitem{CMD-2:2003gqi}
{\bfseries CMD-2} Collaboration, R.~R. Akhmetshin {\em et~al.}, ``{Reanalysis
  of hadronic cross-section measurements at CMD-2},''
  \href{https://dx.doi.org/10.1016/j.physletb.2003.10.108}{{\em Phys. Lett. B}
  {\bfseries 578} (2004) 285--289},
  \href{https://arxiv.org/abs/hep-ex/0308008}{{\ttfamily
  arXiv:hep-ex/0308008}}.

\bibitem{CMD-2:2005mvb}
{\bfseries CMD-2} Collaboration, V.~M. Aul'chenko {\em et~al.}, ``{Measurement
  of the pion form-factor in the range 1.04~GeV to 1.38~GeV with the CMD-2
  detector},'' \href{https://dx.doi.org/10.1134/1.2175241}{{\em JETP Lett.}
  {\bfseries 82} (2005) 743--747},
  \href{https://arxiv.org/abs/hep-ex/0603021}{{\ttfamily
  arXiv:hep-ex/0603021}}.

\bibitem{Aulchenko:2006dxz}
{\bfseries CMD-2} Collaboration, V.~M. Aul'chenko {\em et~al.}, ``{Measurement
  of the $e^+ e^- \to \pi^+ \pi^-$ cross section with the CMD-2 detector in the
  370--520~MeV c.m. energy range},''
  \href{https://dx.doi.org/10.1134/S0021364006200021}{{\em JETP Lett.}
  {\bfseries 84} (2006) 413--417},
  \href{https://arxiv.org/abs/hep-ex/0610016}{{\ttfamily
  arXiv:hep-ex/0610016}}.

\bibitem{CMD-2:2006gxt}
{\bfseries CMD-2} Collaboration, R.~R. Akhmetshin {\em et~al.},
  ``{High-statistics measurement of the pion form factor in the rho-meson
  energy range with the CMD-2 detector},''
  \href{https://dx.doi.org/10.1016/j.physletb.2007.01.073}{{\em Phys. Lett. B}
  {\bfseries 648} (2007) 28--38},
  \href{https://arxiv.org/abs/hep-ex/0610021}{{\ttfamily
  arXiv:hep-ex/0610021}}.

\bibitem{Achasov:2006vp}
{\bfseries SND} Collaboration, M.~N. Achasov {\em et~al.}, ``{Update of the
  $e^+e^- \to \pi^+\pi^-$ cross-section measured by SND detector in the energy
  region $400 < \sqrt{s} < 1000$~MeV},''
  \href{https://dx.doi.org/10.1134/S106377610609007X}{{\em J. Exp. Theor.
  Phys.} {\bfseries 103} (2006) 380--384},
  \href{https://arxiv.org/abs/hep-ex/0605013}{{\ttfamily
  arXiv:hep-ex/0605013}}.

\bibitem{SND:2020nwa}
{\bfseries SND} Collaboration, M.~N. Achasov {\em et~al.}, ``{Measurement of
  the $e^+e^- \to\pi^+\pi^- $ process cross section with the SND detector at
  the VEPP-2000 collider in the energy region $0.525<\sqrt{s}<0.883$ GeV},''
  \href{https://dx.doi.org/10.1007/JHEP01(2021)113}{{\em JHEP} {\bfseries 01}
  (2021) 113}, \href{https://arxiv.org/abs/2004.00263}{{\ttfamily
  arXiv:2004.00263 [hep-ex]}}.

\bibitem{Bryzgalov:2023}
O.~V. Zenin and V.~V. Bryzgalov, ``{Estimation of the hadron contribution to
  $g_\mu-2$ using the IHEP total cross section database},'' in {\em {XXXV
  International Workshop on High Energy Physics ``From Quarks to Galaxies:
  Elucidating Dark Sides'', Protvino, Russia, November 28 -- December 1,
  2023}}.
\newblock \url{{https://indico.ihep.su/event/hepftXXXV}}.

\bibitem{code:2023}
V.~V. Bryzgalov and O.~V. Zenin.
  \url{{https://glab.ihep.su/zenin\_o/compas\_users/-/tree/master/ee/}}.

\bibitem{ParticleDataGroup:2022pth}
{\bfseries Particle Data Group} Collaboration, R.~L. Workman {\em et~al.},
  ``{Review of Particle Physics},''
  \href{https://dx.doi.org/10.1093/ptep/ptac097}{{\em PTEP} {\bfseries 2022}
  (2022) 083C01}.

\bibitem{ParticleDataGroup:2002ivw}
{\bfseries Particle Data Group} Collaboration, K.~Hagiwara {\em et~al.},
  ``{Review of particle physics},''
  \href{https://dx.doi.org/10.1103/PhysRevD.66.010001}{{\em Phys. Rev. D}
  {\bfseries 66} (2002) 010001}.

\bibitem{Ezhela:2003pp}
V.~V. Ezhela, S.~B. Lugovsky, and O.~V. Zenin, ``{Hadronic part of the muon
  $g-2$ estimated on the $\sigma^{2003}_{(tot)}(e^+e^- \to hadrons)$ evaluated
  data compilation},'' IHEP-2003-35,
  \href{https://arxiv.org/abs/hep-ph/0312114}{{\ttfamily
  arXiv:hep-ph/0312114}}.

\bibitem{Kang:2002bm}
{\bfseries COMPETE ($e^+e^-$)} Collaboration, K.~Kang, V.~V. Ezhela, S.~K.
  Kang, S.~B. Lugovsky, M.~R. Whalley, and O.~V. Zenin,
  \href{https://dx.doi.org/10.1142/9789812778352_0018}{``{Total Cross Sections
  of $e^+e^- \to$ hadrons and pQCD Tests},''} in {\em {6th Workshop on
  Non-Perturbative Quantum Chromodynamics}}, pp.~120--133.
\newblock 2002.
\newblock \href{https://arxiv.org/abs/hep-ph/0202051}{{\ttfamily
  arXiv:hep-ph/0202051}}.

\bibitem{Davier:2019can}
M.~Davier, A.~Hoecker, B.~Malaescu, and Z.~Zhang, ``{A new evaluation of the
  hadronic vacuum polarisation contributions to the muon anomalous magnetic
  moment and to $\mathbf{\boldsymbol\alpha(m_Z^2)}$},''
  \href{https://dx.doi.org/10.1140/epjc/s10052-020-7792-2}{{\em Eur. Phys. J.
  C} {\bfseries 80} no.~3, (2020) 241},
  \href{https://arxiv.org/abs/1908.00921}{{\ttfamily arXiv:1908.00921
  [hep-ph]}}. [Erratum: Eur.Phys.J.C 80, 410 (2020)].

\bibitem{Davier:2017zfy}
M.~Davier, A.~Hoecker, B.~Malaescu, and Z.~Zhang, ``{Reevaluation of the
  hadronic vacuum polarisation contributions to the Standard Model predictions
  of the muon $g-2$ and ${\alpha (m_Z^2)}$ using newest hadronic cross-section
  data},'' \href{https://dx.doi.org/10.1140/epjc/s10052-017-5161-6}{{\em Eur.
  Phys. J. C} {\bfseries 77} no.~12, (2017) 827},
  \href{https://arxiv.org/abs/1706.09436}{{\ttfamily arXiv:1706.09436
  [hep-ph]}}.

\bibitem{Keshavarzi:2019abf}
A.~Keshavarzi, D.~Nomura, and T.~Teubner, ``{$g-2$ of charged leptons, $\alpha
  (M^2_Z)$ , and the hyperfine splitting of muonium},''
  \href{https://dx.doi.org/10.1103/PhysRevD.101.014029}{{\em Phys. Rev. D}
  {\bfseries 101} no.~1, (2020) 014029}, MAN/HEP/2019/010, LTH 1216,
  KEK-TH-2165, \href{https://arxiv.org/abs/1911.00367}{{\ttfamily
  arXiv:1911.00367 [hep-ph]}}.

\bibitem{Keshavarzi:2018mgv}
A.~Keshavarzi, D.~Nomura, and T.~Teubner, ``{Muon $g-2$ and $\alpha(M_Z^2)$: a
  new data-based analysis},''
  \href{https://dx.doi.org/10.1103/PhysRevD.97.114025}{{\em Phys. Rev. D}
  {\bfseries 97} no.~11, (2018) 114025}, LTH 1153, KEK-TH-2035, LTH-1153,
  YITP-18-09, LTH 1153; KEK-TH-2035; YITP-18-09,
  \href{https://arxiv.org/abs/1802.02995}{{\ttfamily arXiv:1802.02995
  [hep-ph]}}.

\bibitem{Hoferichter:2019mqg}
M.~Hoferichter, B.-L. Hoid, and B.~Kubis, ``{Three-pion contribution to
  hadronic vacuum polarization},''
  \href{https://dx.doi.org/10.1007/JHEP08(2019)137}{{\em JHEP} {\bfseries 08}
  (2019) 137}, INT-PUB-19-030,
  \href{https://arxiv.org/abs/1907.01556}{{\ttfamily arXiv:1907.01556
  [hep-ph]}}.

\bibitem{Colangelo:2018mtw}
G.~Colangelo, M.~Hoferichter, and P.~Stoffer, ``{Two-pion contribution to
  hadronic vacuum polarization},''
  \href{https://dx.doi.org/10.1007/JHEP02(2019)006}{{\em JHEP} {\bfseries 02}
  (2019) 006}, INT-PUB-18-048,
  \href{https://arxiv.org/abs/1810.00007}{{\ttfamily arXiv:1810.00007
  [hep-ph]}}.

\bibitem{Muong-2:2023cdq}
{\bfseries Muon g-2} Collaboration, D.~P. Aguillard {\em et~al.},
  ``{Measurement of the Positive Muon Anomalous Magnetic Moment to 0.20~ppm},''
  \href{https://dx.doi.org/10.1103/PhysRevLett.131.161802}{{\em Phys. Rev.
  Lett.} {\bfseries 131} no.~16, (2023) 161802},
  FERMILAB-PUB-23-385-AD-CSAID-PPD,
  \href{https://arxiv.org/abs/2308.06230}{{\ttfamily arXiv:2308.06230
  [hep-ex]}}.

\bibitem{BESIII:2015equ}
{\bfseries BES III} Collaboration, M.~Ablikim {\em et~al.}, ``{Measurement of
  the $e^+ e^- \to \pi^+ \pi^-$ cross section between 600 and 900 MeV using
  initial state radiation},''
  \href{https://dx.doi.org/10.1016/j.physletb.2015.11.043}{{\em Phys. Lett. B}
  {\bfseries 753} (2016) 629--638},
  \href{https://arxiv.org/abs/1507.08188}{{\ttfamily arXiv:1507.08188
  [hep-ex]}}. [Erratum: Phys.Lett.B 812, 135982 (2021)].

\bibitem{Bollini:1975nw}
{\bfseries Frascati (ADONE)} Collaboration, D.~Bollini, P.~Giusti, T.~Massam,
  L.~Monari, F.~Palmonari, G.~Valenti, and A.~Zichichi, ``{Search for
  $\rho$-Like Vector Mesons in the Mass Range 1.2~GeV to 3.0~GeV},''
  \href{https://dx.doi.org/10.1007/BF02725904}{{\em Lett. Nuovo Cim.}
  {\bfseries 15} (1976) 393}, PRINT-76-0036 (CERN).

\bibitem{Barkov:1985ac}
{\bfseries CMD} Collaboration, L.~M. Barkov {\em et~al.}, ``{Electromagnetic
  Pion Form-Factor in the Timelike Region},''
  \href{https://dx.doi.org/10.1016/0550-3213(85)90399-2}{{\em Nucl. Phys. B}
  {\bfseries 256} (1985) 365--384}.

\bibitem{DM2:1988xqd}
{\bfseries DM2} Collaboration, D.~Bisello {\em et~al.}, ``{The Pion
  Electromagnetic Form-factor in the Timelike Energy Range 1.35~GeV $\le
  \sqrt{s} \le$ 2.4~GeV},''
  \href{https://dx.doi.org/10.1016/0370-2693(89)90060-9}{{\em Phys. Lett. B}
  {\bfseries 220} (1989) 321--327}, LAL-88-47.

\bibitem{Quenzer:1978qt}
{\bfseries DM1} Collaboration, A.~Quenzer {\em et~al.}, ``{Pion Form-Factor
  from 480~MeV to 1100~MeV},''
  \href{https://dx.doi.org/10.1016/0370-2693(78)90918-8}{{\em Phys. Lett. B}
  {\bfseries 76} (1978) 512--516}.

\bibitem{Xiao:2017dqv}
{\bfseries CLEO-c} Collaboration, T.~Xiao, S.~Dobbs, A.~Tomaradze, K.~K. Seth,
  and G.~Bonvicini, ``{Precision Measurement of the Hadronic Contribution to
  the Muon Anomalous Magnetic Moment},''
  \href{https://dx.doi.org/10.1103/PhysRevD.97.032012}{{\em Phys. Rev. D}
  {\bfseries 97} no.~3, (2018) 032012},
  \href{https://arxiv.org/abs/1712.04530}{{\ttfamily arXiv:1712.04530
  [hep-ex]}}.

\bibitem{Vasserman:1981xq}
{\bfseries TOF} Collaboration, I.~B. Vasserman, P.~M. Ivanov, G.~Y.
  Kezerashvili, I.~A. Koop, A.~P. Lysenko, Y.~N. Pestov, A.~N. Skrinsky, G.~V.
  Fedotovich, and Y.~M. Shatunov, ``{Pion Form-factor Measurement in the
  Reaction $e^+ e^- \to \pi^+ \pi^-$ for Energies Within the Range From 0.4~GeV
  to 0.46~GeV},'' {\em Yad. Fiz.} {\bfseries 33} (1981) 709--714.

\bibitem{Achasov:2002ud}
{\bfseries SND} Collaboration, M.~N. Achasov {\em et~al.}, ``{Study of the
  process $e^+ e^- \to \pi^+ \pi^- \pi^0$ in the energy region $s^{1/2}$ from
  0.98~GeV to 1.38~GeV},''
  \href{https://dx.doi.org/10.1103/PhysRevD.66.032001}{{\em Phys. Rev. D}
  {\bfseries 66} (2002) 032001},
  \href{https://arxiv.org/abs/hep-ex/0201040}{{\ttfamily
  arXiv:hep-ex/0201040}}.

\bibitem{CMD:1989aaa}
{\bfseries CMD} Collaboration, L.~M. Barkov {\em et~al.}, ``{Investigation of
  the Process $e^+ e^- \to \pi^+ \pi^- \pi^0$ at the Energy Range 840 -- 1020
  MeV on VEPP-2M with the Help of Cryogenic Magnetic Detector},'' IYF-89-15.

\bibitem{Cordier:1979qg}
{\bfseries DM1} Collaboration, A.~Cordier {\em et~al.}, ``{Cross-section of the
  Reaction $e^+ e^- \to \pi^+ \pi^- \pi^0$ for Center-of-mass Energies From
  750~MeV to 1100~MeV},''
  \href{https://dx.doi.org/10.1016/0550-3213(80)90157-1}{{\em Nucl. Phys. B}
  {\bfseries 172} (1980) 13--24}, LAL-79/1.

\bibitem{Akhmetshin:1995vz}
{\bfseries CMD-2} Collaboration, R.~R. Akhmetshin {\em et~al.}, ``{Measurement
  of phi meson parameters with CMD-2 detector at VEPP-2M collider},''
  \href{https://dx.doi.org/10.1016/0370-2693(95)01394-6}{{\em Phys. Lett. B}
  {\bfseries 364} (1995) 199--206}.

\bibitem{Akhmetshin:1998se}
{\bfseries CMD-2} Collaboration, R.~R. Akhmetshin {\em et~al.}, ``{Study of
  dynamics of $\phi \to \pi^+ \pi^- \pi^0$ decay with CMD-2 detector},''
  \href{https://dx.doi.org/10.1016/S0370-2693(98)00826-0}{{\em Phys. Lett. B}
  {\bfseries 434} (1998) 426--436}.

\bibitem{Achasov:2000am}
{\bfseries SND} Collaboration, M.~N. Achasov {\em et~al.}, ``{Measurements of
  the parameters of the $\phi(1020)$ resonance through studies of the processes
  $e^+ e^- \to K^+ K^-$, $K_SK_L$, and $\pi^+ \pi^- \pi^0$},''
  \href{https://dx.doi.org/10.1103/PhysRevD.63.072002}{{\em Phys. Rev. D}
  {\bfseries 63} (2001) 072002},
  \href{https://arxiv.org/abs/hep-ex/0009036}{{\ttfamily
  arXiv:hep-ex/0009036}}.

\bibitem{Achasov:2003ir}
{\bfseries SND} Collaboration, M.~N. Achasov {\em et~al.}, ``{Study of the
  process $e^+ e^- \to \pi^+ \pi^- \pi^0$ in the energy region $\sqrt{s} <
  0.98$~GeV},'' \href{https://dx.doi.org/10.1103/PhysRevD.68.052006}{{\em Phys.
  Rev. D} {\bfseries 68} (2003) 052006},
  \href{https://arxiv.org/abs/hep-ex/0305049}{{\ttfamily
  arXiv:hep-ex/0305049}}.

\bibitem{BaBar:2004ytv}
{\bfseries BaBar} Collaboration, B.~Aubert {\em et~al.}, ``{Study of $e^+e^-
  \to \pi^+ \pi^- \pi^0$ process using initial state radiation with BaBar},''
  \href{https://dx.doi.org/10.1103/PhysRevD.70.072004}{{\em Phys. Rev. D}
  {\bfseries 70} (2004) 072004}, SLAC-PUB-10624, BABAR-PUB-04-034,
  \href{https://arxiv.org/abs/hep-ex/0408078}{{\ttfamily
  arXiv:hep-ex/0408078}}.

\bibitem{Dolinsky:1991vq}
{\bfseries ND} Collaboration, S.~I. Dolinsky {\em et~al.}, ``{Summary of
  experiments with the neutral detector at the $e^+ e^-$ storage ring
  VEPP-2M},'' \href{https://dx.doi.org/10.1016/0370-1573(91)90127-8}{{\em Phys.
  Rept.} {\bfseries 202} (1991) 99--170}.

\bibitem{Aulchenko:2015mwt}
{\bfseries CMD-2} Collaboration, V.~M. Aul'chenko {\em et~al.}, ``{Study of the
  e$^{+}$ e$^{–}$ \textrightarrow{}
  \ensuremath{\pi}$^{+}$\ensuremath{\pi}$^{–}$\ensuremath{\pi}$^{0}$ process
  in the energy range 1.05\textendash{}2.00 GeV},''
  \href{https://dx.doi.org/10.1134/S1063776115060023}{{\em J. Exp. Theor.
  Phys.} {\bfseries 121} no.~1, (2015) 27--34}.

\bibitem{DM2:1992zkc}
{\bfseries DM2} Collaboration, A.~Antonelli {\em et~al.}, ``{Measurement of the
  $e^+ e^- \to \pi^+ \pi^- \pi^0$ and $e^+ e^- \to \omega \pi^+ \pi^-$
  reactions in the energy interval 1350~MeV--2400~MeV},''
  \href{https://dx.doi.org/10.1007/BF01589702}{{\em Z. Phys. C} {\bfseries 56}
  (1992) 15--20}, LAL-92-08.

\bibitem{SND:2001mhi}
{\bfseries SND} Collaboration, M.~N. Achasov {\em et~al.}, ``{$e^+ e^- \to
  4\pi$ processes investigation in the energy range 0.98~GeV to 1.38~GeV with
  SND detector},'' BUDKER-INP-2001-34.

\bibitem{DM2:1990kxb}
{\bfseries DM2} Collaboration, D.~Bisello {\em et~al.}, ``{DM2 results on $e^+
  e^-$ annihilation into multi - hadrons in the 1350~MeV - 2400~MeV energy
  range},'' in {\em {25th International Conference on High-energy Physics}}.
\newblock 6, 1990.

\bibitem{Esposito:1981dv}
{\bfseries MEA} Collaboration, B.~Esposito {\em et~al.}, ``{Measurement on
  $\pi^+ \pi^- \pi^0 \pi^0$, $\pi^+ \pi^- \pi^+ \pi^- \pi^0$, $\pi^+ \pi^-
  \pi^+ \pi^- \pi^0 \pi^0$, $\pi^+ \pi^- \pi^+ \pi^- \pi^+ \pi^-$ Production
  Cross-sections in $e^+ e^-$ Annihilation at 1.45~GeV - 1.80~GeV
  Center-of-mass Energy},'' \href{https://dx.doi.org/10.1007/BF02776174}{{\em
  Lett. Nuovo Cim.} {\bfseries 31} (1981) 445--452}.

\bibitem{Cosme:1978qe}
{\bfseries M3N} Collaboration, G.~Cosme {\em et~al.}, ``{Hadronic
  Cross-sections Study in $e^+ e^-$ Collisions From 1.350~GeV to 2.125~GeV},''
  \href{https://dx.doi.org/10.1016/0550-3213(79)90100-7}{{\em Nucl. Phys. B}
  {\bfseries 152} (1979) 215--231}, LAL-78-32.

\bibitem{Bacci:1980zs}
{\bfseries $\gamma\gamma{}2$} Collaboration, C.~Bacci {\em et~al.},
  ``{Measurement of Hadronic Exclusive Cross-sections in $e^+ e^-$ Annihilation
  From 1.42~GeV to 2.20~GeV},''
  \href{https://dx.doi.org/10.1016/0550-3213(81)90208-X}{{\em Nucl. Phys. B}
  {\bfseries 184} (1981) 31--39}, LNF-80/72-P.

\bibitem{CMD-2:1998gab}
{\bfseries CMD-2} Collaboration, R.~R. Akhmetshin {\em et~al.}, ``{$a(1)(1260)
  \pi$ dominance in the process $e^+ e^- \to 4\pi$ at energies 1.05~GeV --
  1.38~GeV},'' \href{https://dx.doi.org/10.1016/S0370-2693(99)01080-1}{{\em
  Phys. Lett. B} {\bfseries 466} (1999) 392--402}, BUDKER-INP-1998-83,
  \href{https://arxiv.org/abs/hep-ex/9904024}{{\ttfamily
  arXiv:hep-ex/9904024}}.

\bibitem{Cosme:1976tf}
{\bfseries ACO} Collaboration, G.~Cosme {\em et~al.}, ``{Multi-Pion Production
  Below 1.1~GeV by $e^+ e^-$ Annihilation},''
  \href{https://dx.doi.org/10.1016/0370-2693(76)90280-X}{{\em Phys. Lett. B}
  {\bfseries 63} (1976) 349--351}.

\bibitem{BaBar:2017zmc}
{\bfseries BaBar} Collaboration, J.~P. Lees {\em et~al.}, ``{Measurement of the
  ${e}^{+}{e}^{{-}}{\rightarrow}{{\pi}}^{+}{{\pi}}^{{-}}{{\pi}}^{0}{{\pi}}^{0}$
  cross section using initial-state radiation at BABAR},''
  \href{https://dx.doi.org/10.1103/PhysRevD.96.092009}{{\em Phys. Rev. D}
  {\bfseries 96} no.~9, (2017) 092009}, SLAC-PUB-17147, BABAR-PUB-17-002,
  \href{https://arxiv.org/abs/1709.01171}{{\ttfamily arXiv:1709.01171
  [hep-ex]}}.

\bibitem{Kurdadze:1986tc}
{\bfseries OLYA} Collaboration, L.~M. Kurdadze {\em et~al.}, ``{Study of the
  Reaction $e^+ e^- \to \pi^+ \pi^- \pi^0 \pi^0$ at (2 $e$) Up to 1.4~GeV},''
  {\em JETP Lett.} {\bfseries 43} (1986) 643--645.

\bibitem{Cordier:1981zn}
{\bfseries DM1} Collaboration, A.~Cordier, D.~Bisello, J.~C. Bizot, J.~Buon,
  B.~Delcourt, L.~Fayard, and F.~Mane, ``{Study of the $e^+ e^- \to \pi^+ \pi^-
  \pi^+ \pi^-$ Reaction in the 1.4~GeV to 2.18~GeV Energy Range},''
  \href{https://dx.doi.org/10.1016/0370-2693(82)90478-6}{{\em Phys. Lett. B}
  {\bfseries 109} (1982) 129--132}, LAL 81/33.

\bibitem{CMD-2:1999dof}
{\bfseries CMD-2} Collaboration, R.~R. Akhmetshin {\em et~al.},
  ``{Cross-section of the reaction $e^+ e^- \to \pi^+ \pi^- \pi^+ \pi^-$ below
  1~GeV at CMD-2},''
  \href{https://dx.doi.org/10.1016/S0370-2693(00)00043-5}{{\em Phys. Lett. B}
  {\bfseries 475} (2000) 190--197},
  \href{https://arxiv.org/abs/hep-ex/9912020}{{\ttfamily
  arXiv:hep-ex/9912020}}.

\bibitem{CMD-2:2000rfr}
{\bfseries CMD-2} Collaboration, R.~R. Akhmetshin {\em et~al.}, ``{Observation
  of the $\phi \to \pi^+ \pi^- \pi^+ \pi^-$ decay},''
  \href{https://dx.doi.org/10.1016/S0370-2693(00)01001-7}{{\em Phys. Lett. B}
  {\bfseries 491} (2000) 81--89},
  \href{https://arxiv.org/abs/hep-ex/0008019}{{\ttfamily
  arXiv:hep-ex/0008019}}.

\bibitem{CMD-2:2004vyr}
{\bfseries CMD-2} Collaboration, R.~R. Akhmetshin {\em et~al.}, ``{Total cross
  section of the process $e^+ e^- \to \pi^+ \pi^- \pi^+ \pi^-$ in the C.M.
  energy range 980~MeV to 1380~MeV},''
  \href{https://dx.doi.org/10.1016/j.physletb.2004.05.056}{{\em Phys. Lett. B}
  {\bfseries 595} (2004) 101--108},
  \href{https://arxiv.org/abs/hep-ex/0404019}{{\ttfamily
  arXiv:hep-ex/0404019}}.

\bibitem{Akhmetshin:2016dtr}
{\bfseries CMD-3} Collaboration, R.~R. Akhmetshin {\em et~al.}, ``{Study of the
  process $e^+e^-\to \pi^+\pi^-\pi^+\pi^-$ in the c.m. energy range
  920\textendash{}1060 MeV with the CMD-3 detector},''
  \href{https://dx.doi.org/10.1016/j.physletb.2017.03.022}{{\em Phys. Lett. B}
  {\bfseries 768} (2017) 345--350},
  \href{https://arxiv.org/abs/1612.04483}{{\ttfamily arXiv:1612.04483
  [hep-ex]}}.

\bibitem{Cordier:1978yp}
{\bfseries DM1} Collaboration, A.~Cordier {\em et~al.}, ``{Cross-section of the
  Reaction $e^+ e^- \to \pi^+ \pi^- \pi^+ \pi^-$ for Center-of-mass Energies
  From 890~MeV to 1100~MeV},''
  \href{https://dx.doi.org/10.1016/0370-2693(79)90360-5}{{\em Phys. Lett. B}
  {\bfseries 81} (1979) 389--392}, LAL-78/31.

\bibitem{Bacci:1980ru}
{\bfseries $\gamma\gamma{}2$} Collaboration, C.~Bacci {\em et~al.},
  ``{Measurement of the $e^+ e^- \to \pi^+ \pi^- \pi^+ \pi^-$ Cross-section in
  the $\rho^\prime$ (1600) Energy Region},''
  \href{https://dx.doi.org/10.1016/0370-2693(80)90418-9}{{\em Phys. Lett. B}
  {\bfseries 95} (1980) 139--142}, LNF-80/25-P.

\bibitem{BaBar:2012sxt}
{\bfseries BaBar} Collaboration, J.~P. Lees {\em et~al.}, ``{Initial-State
  Radiation Measurement of the $e^+e^- -> \pi^+\pi^-\pi^+\pi^-$ Cross
  Section},'' \href{https://dx.doi.org/10.1103/PhysRevD.85.112009}{{\em Phys.
  Rev. D} {\bfseries 85} (2012) 112009}, SLAC-PUB-14857, BABAR-PUB-11-016,
  \href{https://arxiv.org/abs/1201.5677}{{\ttfamily arXiv:1201.5677 [hep-ex]}}.

\bibitem{Barkov:1988gp}
{\bfseries CMD} Collaboration, L.~M. Barkov {\em et~al.}, ``{The investigation
  of multi - pion creation with the cryogenic magnetic detector at the VEPP-2M
  storage ring. (In Russian)},'' {\em Sov. J. Nucl. Phys.} {\bfseries 47}
  (1988) 248--252.

\bibitem{Kurdadze:1988mu}
{\bfseries OLYA} Collaboration, L.~M. Kurdadze, M.~Y. Lelchuk, E.~V.
  Pakhtusova, V.~A. Sidorov, A.~N. Skrinsky, A.~G. Chilingarov, Y.~M. Shatunov,
  B.~A. Shvarts, and S.~I. Eidelman, ``{Study of $e^+ e^- \to \pi^+ \pi^- \pi^+
  \pi^-$ Reaction at 2 $e$ Up to 1.4~GeV},'' {\em JETP Lett.} {\bfseries 47}
  (1988) 512--515.

\bibitem{CMD-2:2008fsu}
{\bfseries CMD-2} Collaboration, R.~R. Akhmetshin {\em et~al.}, ``{Measurement
  of $e^+e^- \to \phi \to K^+K^-$ cross section with the CMD-2 detector at
  VEPP-2M Collider},''
  \href{https://dx.doi.org/10.1016/j.physletb.2008.09.053}{{\em Phys. Lett. B}
  {\bfseries 669} (2008) 217--222}, 0804.0146,
  \href{https://arxiv.org/abs/0804.0178}{{\ttfamily arXiv:0804.0178 [hep-ex]}}.

\bibitem{Kozyrev:2017agm}
{\bfseries CMD-3} Collaboration, E.~A. Kozyrev {\em et~al.}, ``{Study of the
  process $e^+e^- \to K^+K^-$ in the center-of-mass energy range
  1010--1060\textasciitilde{}MeV with the CMD-3 detector},''
  \href{https://dx.doi.org/10.1016/j.physletb.2018.01.079}{{\em Phys. Lett. B}
  {\bfseries 779} (2018) 64--71},
  \href{https://arxiv.org/abs/1710.02989}{{\ttfamily arXiv:1710.02989
  [hep-ex]}}.

\bibitem{CMD:1983aaa}
{\bfseries CMD} Collaboration, G.~V. Anikin {\em et~al.}, ``{The Results of
  Experiments with CMD on VEPP-2M Storage Ring},'' in {\em {Ithaca 1983,
  International Symposium on Lepton and Photon Interactions at High Energies,
  Ithaca, N.Y., Aug 4-9, 1983}}.
\newblock 8, 1983.

\bibitem{Esposito:1980bz}
{\bfseries MEA} Collaboration, B.~Esposito {\em et~al.}, ``{Measurements of the
  EM timelike form-factors for kaon and pion at $\sqrt{s} = 1.5$~GeV},''
  \href{https://dx.doi.org/10.1007/BF02804624}{{\em Lett. Nuovo Cim.}
  {\bfseries 28} (1980) 337--342}.

\bibitem{Ivanov:1981wf}
{\bfseries OLYA} Collaboration, P.~M. Ivanov {\em et~al.}, ``{Measurement of
  the Charged Kaon Form-factor in the Energy Range 1.0~GeV to 1.4~GeV},''
  \href{https://dx.doi.org/10.1016/0370-2693(81)90834-0}{{\em Phys. Lett. B}
  {\bfseries 107} (1981) 297--300}.

\bibitem{Delcourt:1980eq}
{\bfseries DM1} Collaboration, B.~Delcourt, D.~Bisello, J.~C. Bizot, J.~Buon,
  A.~Cordier, and F.~Mane, ``{Study of the Reaction $e^+ e^- \to K^+ K^-$ in
  the Total Energy Range 1400~MeV to 2060~MeV},''
  \href{https://dx.doi.org/10.1016/0370-2693(81)91121-7}{{\em Phys. Lett. B}
  {\bfseries 99} (1981) 257--260}, LAL/80-36.

\bibitem{BaBar:2013jqz}
{\bfseries BaBar} Collaboration, J.~P. Lees {\em et~al.}, ``{Precision
  measurement of the $e^+e^- \to K^+K^-(\gamma)$ cross section with the
  initial-state radiation method at BABAR},''
  \href{https://dx.doi.org/10.1103/PhysRevD.88.032013}{{\em Phys. Rev. D}
  {\bfseries 88} no.~3, (2013) 032013}, BABAR-PUB-13-006, SLAC-PUB-15487,
  \href{https://arxiv.org/abs/1306.3600}{{\ttfamily arXiv:1306.3600 [hep-ex]}}.

\bibitem{Achasov:2016lbc}
{\bfseries SND} Collaboration, M.~N. Achasov {\em et~al.}, ``{Measurement of
  the $\mathbf{e^+e^-\to K^+K^-}$ cross section in the energy range
  $\mathbf{\sqrt{s}=1.05-2.0}$ GeV}''
  \href{https://dx.doi.org/10.1103/PhysRevD.94.112006}{{\em Phys. Rev. D}
  {\bfseries 94} no.~11, (2016) 112006},
  \href{https://arxiv.org/abs/1608.08757}{{\ttfamily arXiv:1608.08757
  [hep-ex]}}.

\bibitem{CMD-2:1999chh}
{\bfseries CMD-2} Collaboration, R.~R. Akhmetshin {\em et~al.}, ``{Measurement
  of $\phi$ meson parameters in K$^0_L$ K$^0_S$ decay mode with CMD-2},''
  \href{https://dx.doi.org/10.1016/S0370-2693(99)00973-9}{{\em Phys. Lett. B}
  {\bfseries 466} (1999) 385},
  \href{https://arxiv.org/abs/hep-ex/9906032}{{\ttfamily
  arXiv:hep-ex/9906032}}. [Erratum: Phys.Lett.B 508, 217--218 (2001)].

\bibitem{Akhmetshin:2002vj}
{\bfseries CMD-2} Collaboration, R.~R. Akhmetshin {\em et~al.}, ``{Study of the
  process $e^+ e^- \to K^0_L K^0_S$ in the CM energy range 1.05~GeV to 1.38~GeV
  with CMD-2},'' \href{https://dx.doi.org/10.1016/S0370-2693(02)02985-4}{{\em
  Phys. Lett. B} {\bfseries 551} (2003) 27--34},
  \href{https://arxiv.org/abs/hep-ex/0211004}{{\ttfamily
  arXiv:hep-ex/0211004}}.

\bibitem{CMD-3:2016nhy}
{\bfseries CMD-3} Collaboration, E.~A. Kozyrev {\em et~al.}, ``{Study of the
  process $e^+ e^- \to K^0_{S}K^0_{L}$ in the center-of-mass energy range
  1004--1060 MeV with the CMD-3 detector at the VEPP-2000 $e^+ e^-$
  collider},'' \href{https://dx.doi.org/10.1016/j.physletb.2016.07.003}{{\em
  Phys. Lett. B} {\bfseries 760} (2016) 314--319},
  \href{https://arxiv.org/abs/1604.02981}{{\ttfamily arXiv:1604.02981
  [hep-ex]}}.

\bibitem{Mane:1980ep}
{\bfseries DM1} Collaboration, F.~Mane, D.~Bisello, J.~C. Bizot, J.~Buon,
  A.~Cordier, and B.~Delcourt, ``{Study of the Reaction $e^+ e^- \to K^0_S
  K^0_L$ in the Total Energy Range 1.4~GeV to 2.18~GeV and Interpretation of
  the $K^+$ and $K^0$ Form-factors},''
  \href{https://dx.doi.org/10.1016/0370-2693(81)91122-9}{{\em Phys. Lett. B}
  {\bfseries 99} (1981) 261--264}, LAL/80-37.

\bibitem{BaBar:2014uwz}
{\bfseries BaBar} Collaboration, J.~P. Lees {\em et~al.}, ``{Cross sections for
  the reactions $e^+ e^-\to K_S^0 K_L^0$, $K_S^0 K_L^0 \pi^+\pi^-$, $K_S^0
  K_S^0 \pi^+\pi^-$, and $K_S^0 K_S^0 K^+K^-$ from events with initial-state
  radiation},'' \href{https://dx.doi.org/10.1103/PhysRevD.89.092002}{{\em Phys.
  Rev. D} {\bfseries 89} no.~9, (2014) 092002}, SLAC-PUB-15934,
  \href{https://arxiv.org/abs/1403.7593}{{\ttfamily arXiv:1403.7593 [hep-ex]}}.

\bibitem{Ivanov:1982cr}
{\bfseries OLYA} Collaboration, P.~M. Ivanov {\em et~al.}, ``{Measurements of
  the form-factor of the neutral kaon from 1.06~GeV to 1.40~GeV},'' {\em JETP
  Lett.} {\bfseries 36} (1982) 112--115.

\bibitem{Achasov:2000zd}
{\bfseries SND} Collaboration, M.~N. Achasov {\em et~al.}, ``{Experimental
  study of the processes $e^+ e^- \to \phi \to \eta \gamma$, $\pi^0 \gamma$ at
  VEPP-2M},'' \href{https://dx.doi.org/10.1007/s100529900222}{{\em Eur. Phys.
  J. C} {\bfseries 12} (2000) 25--33}, BUDKER-INP-1999-39.

\bibitem{CMD-2:2004ahv}
{\bfseries CMD-2} Collaboration, R.~R. Akhmetshin {\em et~al.}, ``{Study of the
  processes $e^+ e^- \to \eta \gamma$, $\pi^0 \gamma \to 3\gamma$ in the c.m.
  energy range 600~MeV to 1380~MeV at CMD-2},''
  \href{https://dx.doi.org/10.1016/j.physletb.2004.11.020}{{\em Phys. Lett. B}
  {\bfseries 605} (2005) 26--36}, BUDKER-INP-2004-51,
  \href{https://arxiv.org/abs/hep-ex/0409030}{{\ttfamily
  arXiv:hep-ex/0409030}}.

\bibitem{SND:2016drm}
{\bfseries SND} Collaboration, M.~N. Achasov {\em et~al.}, ``{Study of the
  reaction $e^+e^- \to \pi^0\gamma$ with the SND detector at the VEPP-2M
  collider},'' \href{https://dx.doi.org/10.1103/PhysRevD.93.092001}{{\em Phys.
  Rev. D} {\bfseries 93} no.~9, (2016) 092001},
  \href{https://arxiv.org/abs/1601.08061}{{\ttfamily arXiv:1601.08061
  [hep-ex]}}.

\bibitem{BaBar:2007ceh}
{\bfseries BaBar} Collaboration, B.~Aubert {\em et~al.}, ``{Measurements of
  $e^{+} e^{-} \to K^{+} K^{-} \eta$, $K^{+} K^{-} \pi^0$ and $K^0_{s} K^\pm
  \pi^\mp$ cross- sections using initial state radiation events},''
  \href{https://dx.doi.org/10.1103/PhysRevD.77.092002}{{\em Phys. Rev. D}
  {\bfseries 77} (2008) 092002}, SLAC-PUB-12968, BABAR-PUB-07-052,
  \href{https://arxiv.org/abs/0710.4451}{{\ttfamily arXiv:0710.4451 [hep-ex]}}.

\bibitem{BaBar:2007qju}
{\bfseries BaBar} Collaboration, B.~Aubert {\em et~al.}, ``{The $e^+ e^- \to
  2(\pi^+ \pi^-) \pi^0$, $2(\pi^+ \pi^-) \eta$, $K^+ K^- \pi^+ \pi^- \pi^0$ and
  $K^+ K^- \pi^+ \pi^- \eta$ Cross Sections Measured with Initial-State
  Radiation},'' \href{https://dx.doi.org/10.1103/PhysRevD.76.092005}{{\em Phys.
  Rev. D} {\bfseries 76} (2007) 092005}, BABAR-PUB-07-045, SLAC-PUB-12753,
  \href{https://arxiv.org/abs/0708.2461}{{\ttfamily arXiv:0708.2461 [hep-ex]}}.
  [Erratum: Phys.Rev.D 77, 119902 (2008)].

\bibitem{BaBar:2006vzy}
{\bfseries BaBar} Collaboration, B.~Aubert {\em et~al.}, ``{The $e^+e^- \to
  3(\pi^+ \pi^-), 2(\pi^+ \pi^- \pi^0)$ and $K^+ K^- 2(\pi^+ \pi^-)$ cross
  sections at center-of-mass energies from production threshold to 4.5~GeV
  measured with initial-state radiation},''
  \href{https://dx.doi.org/10.1103/PhysRevD.73.052003}{{\em Phys. Rev. D}
  {\bfseries 73} (2006) 052003}, SLAC-PUB-11663, BABAR-PUB-05-053,
  \href{https://arxiv.org/abs/hep-ex/0602006}{{\ttfamily
  arXiv:hep-ex/0602006}}.

\bibitem{Schioppa:1986aaa}
{\bfseries DM2} Collaboration, M.~Schioppa , ROMA-THESIS-1986-SCHIOPPA.

\bibitem{BaBar:2021rki}
{\bfseries BaBar} Collaboration, J.~P. Lees {\em et~al.}, ``{Study of the
  reactions $e^+e^-\to2(\pi^+\pi^-)\pi^0\pi^0\pi^0$ and
  $e^+e^-\to2(\pi^+\pi^-)\pi^0\pi^0\eta$ at center-of-mass energies from
  threshold to 4.5 GeV using initial-state radiation},''
  \href{https://dx.doi.org/10.1103/PhysRevD.103.092001}{{\em Phys. Rev. D}
  {\bfseries 103} no.~9, (2021) 092001}, SLAC-PUB-17587, BaBar-PUB-20004,
  \href{https://arxiv.org/abs/2102.01314}{{\ttfamily arXiv:2102.01314
  [hep-ex]}}.

\bibitem{Bisello:1981sh}
{\bfseries DM1} Collaboration, D.~Bisello, J.~C. Bizot, J.~Buon, A.~Cordier,
  B.~Delcourt, and F.~Mane, ``{Study of the Reaction $e^+ e^- \to 3 \pi^+ 3
  \pi^-$ in the Total Energy Range 1400~MeV to 2180~MeV},''
  \href{https://dx.doi.org/10.1016/0370-2693(81)91169-2}{{\em Phys. Lett. B}
  {\bfseries 107} (1981) 145--147}, LAL 81/19.

\bibitem{CMD-3:2013nph}
{\bfseries CMD-3} Collaboration, R.~R. Akhmetshin {\em et~al.}, ``{Study of the
  process $e^+e^-\to 3(\pi^+\pi^-)$ in the c.m. energy range 1.5--2.0~GeV with
  the CMD-3 detector},''
  \href{https://dx.doi.org/10.1016/j.physletb.2013.04.065}{{\em Phys. Lett. B}
  {\bfseries 723} (2013) 82--89},
  \href{https://arxiv.org/abs/1302.0053}{{\ttfamily arXiv:1302.0053 [hep-ex]}}.

\bibitem{CMD-3:2019ufp}
{\bfseries CMD-3} Collaboration, R.~R. Akhmetshin {\em et~al.}, ``{Study of the
  process $e^+e^-\to 3(\pi^+\pi^-)\pi^0$ in the C.M. Energy range 1.6--2.0 GeV
  with the CMD-3 detector},''
  \href{https://dx.doi.org/10.1016/j.physletb.2019.04.007}{{\em Phys. Lett. B}
  {\bfseries 792} (2019) 419--423},
  \href{https://arxiv.org/abs/1902.06449}{{\ttfamily arXiv:1902.06449
  [hep-ex]}}.

\bibitem{CMD-2:2001dnv}
{\bfseries CMD-2} Collaboration, R.~R. Akhmetshin {\em et~al.}, ``{Study of the
  process $e^+ e^- \to \eta \gamma$ in center-of-mass energy range 600~MeV to
  1380~MeV at CMD-2},''
  \href{https://dx.doi.org/10.1016/S0370-2693(01)00567-6}{{\em Phys. Lett. B}
  {\bfseries 509} (2001) 217--226},
  \href{https://arxiv.org/abs/hep-ex/0103043}{{\ttfamily
  arXiv:hep-ex/0103043}}.

\bibitem{Cosme:1975rs}
{\bfseries ACO} Collaboration, G.~Cosme {\em et~al.}, ``{New Measurements with
  the Orsay Electron-Positron Storage Ring of the Radiative Decay Modes of the
  $\phi$-Meson},'' \href{https://dx.doi.org/10.1016/0370-2693(76)90281-1}{{\em
  Phys. Lett. B} {\bfseries 63} (1976) 352--356}, LAL-1279.

\bibitem{CMD-2:2000mlo}
{\bfseries CMD-2} Collaboration, R.~R. Akhmetshin {\em et~al.}, ``{Study of the
  process $e^+ e^- \to \pi^+ \pi^- \pi^+ \pi^- \pi^0$ with CMD-2 detector},''
  \href{https://dx.doi.org/10.1016/S0370-2693(00)00937-0}{{\em Phys. Lett. B}
  {\bfseries 489} (2000) 125--130},
  \href{https://arxiv.org/abs/hep-ex/0009013}{{\ttfamily
  arXiv:hep-ex/0009013}}.

\bibitem{SND:2014rfi}
{\bfseries SND} Collaboration, V.~M. Aulchenko {\em et~al.}, ``{Measurement of
  the $e^+e^- \to \eta\pi^+\pi^-$ cross section in the center-of-mass energy
  range 1.22-2.00 GeV with the SND detector at the VEPP-2000 collider},''
  \href{https://dx.doi.org/10.1103/PhysRevD.91.052013}{{\em Phys. Rev. D}
  {\bfseries 91} no.~5, (2015) 052013},
  \href{https://arxiv.org/abs/1412.1971}{{\ttfamily arXiv:1412.1971 [hep-ex]}}.

\bibitem{BaBar:2018erh}
{\bfseries BaBar} Collaboration, J.~P. Lees {\em et~al.}, ``{Study of the
  process $e^+e^- \to \pi^+\pi^-\eta $ using initial state radiation},''
  \href{https://dx.doi.org/10.1103/PhysRevD.97.052007}{{\em Phys. Rev. D}
  {\bfseries 97} (2018) 052007}, BABAR-PUB-17-003, SLAC-PUB-17204,
  \href{https://arxiv.org/abs/1801.02960}{{\ttfamily arXiv:1801.02960
  [hep-ex]}}.

\bibitem{DM2:1990npw}
{\bfseries DM2} Collaboration, D.~Bisello {\em et~al.}, ``{$e^+ e^-$
  annihilation into multi-hadrons in the 1350~MeV - 2400~MeV energy range},''
  \href{https://dx.doi.org/10.1016/0920-5632(91)90244-9}{{\em Nucl. Phys. B
  Proc. Suppl.} {\bfseries 21} (1991) 111--117}, LAL-90-71.

\bibitem{Bisello:1991kd}
{\bfseries DM2} Collaboration, D.~Bisello {\em et~al.}, ``{Observation of an
  isoscalar vector meson at approximately 1650~MeV/$c^2$ in the $e^+ e^- \to K
  \bar{K} \pi$ reaction},'' \href{https://dx.doi.org/10.1007/BF01560440}{{\em
  Z. Phys. C} {\bfseries 52} (1991) 227--230}, LAL-91-24.

\bibitem{BaBar:2011btv}
{\bfseries BaBar} Collaboration, J.~P. Lees {\em et~al.}, ``{Cross Sections for
  the Reactions $e^+ e^- \to K^+ K^- \pi^+ \pi^-$, $K^+ K^- \pi^0 \pi^0$, and
  $K^+ K^- K^+ K^-$ Measured Using Initial-State Radiation Events},''
  \href{https://dx.doi.org/10.1103/PhysRevD.86.012008}{{\em Phys. Rev. D}
  {\bfseries 86} (2012) 012008}, SLAC-PUB-14403, BABAR-PUB-11-001,
  \href{https://arxiv.org/abs/1103.3001}{{\ttfamily arXiv:1103.3001 [hep-ex]}}.

\bibitem{Cordier:1981en}
{\bfseries DM1} Collaboration, A.~Cordier, D.~Bisello, J.~C. Bizot, J.~Buon,
  B.~Delcourt, L.~Fayard, and F.~Mane, ``{Study of the $e^+ e^- \to \pi^+ \pi^-
  K^+ K^-$ Reaction From 1.4~GeV to 2.18~GeV},''
  \href{https://dx.doi.org/10.1016/0370-2693(82)91267-9}{{\em Phys. Lett. B}
  {\bfseries 110} (1982) 335--338}, LAL-81/35.

\bibitem{Shemyakin:2015cba}
{\bfseries CMD-3} Collaboration, D.~N. Shemyakin {\em et~al.}, ``{Measurement
  of the $e^+e^- \to K^+K^-\pi^+\pi^-$ cross section with the CMD-3 detector at
  the VEPP-2000 collider},''
  \href{https://dx.doi.org/10.1016/j.physletb.2016.02.072}{{\em Phys. Lett. B}
  {\bfseries 756} (2016) 153--160},
  \href{https://arxiv.org/abs/1510.00654}{{\ttfamily arXiv:1510.00654
  [hep-ex]}}.

\bibitem{BaBar:2017nrz}
{\bfseries BaBar} Collaboration, J.~P. Lees {\em et~al.}, ``{Cross sections for
  the reactions $e^+ e^-\to K^0_S K^0_L\pi^0$, $K^0_S K^0_L\eta$, and $K^0_S
  K^0_L\pi^0\pi^0$ from events with initial-state radiation},''
  \href{https://dx.doi.org/10.1103/PhysRevD.95.052001}{{\em Phys. Rev. D}
  {\bfseries 95} no.~5, (2017) 052001}, BABAR-PUB-16-004, SLAC-PUB-16918,
  \href{https://arxiv.org/abs/1701.08297}{{\ttfamily arXiv:1701.08297
  [hep-ex]}}.

\bibitem{BaBar:2017pkz}
{\bfseries BaBar} Collaboration, J.~P. Lees {\em et~al.}, ``{Measurement of the
  $e^+e^-\to K^0_{\scriptscriptstyle S}K^\pm\pi^{\mp}\pi^0$ and
  $K^0_{\scriptscriptstyle S}K^\pm\pi^\mp\eta$ cross sections using
  initial-state radiation},''
  \href{https://dx.doi.org/10.1103/PhysRevD.95.092005}{{\em Phys. Rev. D}
  {\bfseries 95} no.~9, (2017) 092005}, BABAR-PUB-15-005, SLAC-PUB-16940,
  \href{https://arxiv.org/abs/1704.05009}{{\ttfamily arXiv:1704.05009
  [hep-ex]}}.

\bibitem{CMD-3:2017tgb}
{\bfseries CMD-3} Collaboration, R.~R. Akhmetshin {\em et~al.}, ``{Study of the
  process $e^+e^-\to \pi^+\pi^-\pi^0\eta$ in the c.m. energy range 1394-2005
  MeV with the CMD-3 detector},''
  \href{https://dx.doi.org/10.1016/j.physletb.2017.08.019}{{\em Phys. Lett. B}
  {\bfseries 773} (2017) 150--158},
  \href{https://arxiv.org/abs/1706.06267}{{\ttfamily arXiv:1706.06267
  [hep-ex]}}.

\bibitem{Achasov:2016qvd}
{\bfseries SND} Collaboration, M.~N. Achasov {\em et~al.}, ``{Measurement of
  the $e^+e^- \to \omega\eta$ cross section below $\sqrt{s}=2$ GeV},''
  \href{https://dx.doi.org/10.1103/PhysRevD.94.092002}{{\em Phys. Rev. D}
  {\bfseries 94} no.~9, (2016) 092002},
  \href{https://arxiv.org/abs/1607.00371}{{\ttfamily arXiv:1607.00371
  [hep-ex]}}.

\bibitem{KLOE:2008woc}
{\bfseries KLOE} Collaboration, F.~Ambrosino {\em et~al.}, ``{Study of the
  process $e^+ e^- \to \omega \pi^0$ in the $\phi$-meson mass region with the
  KLOE detector},''
  \href{https://dx.doi.org/10.1016/j.physletb.2008.09.056}{{\em Phys. Lett. B}
  {\bfseries 669} (2008) 223--228},
  \href{https://arxiv.org/abs/0807.4909}{{\ttfamily arXiv:0807.4909 [hep-ex]}}.

\bibitem{Aulchenko:2000mr}
{\bfseries SND} Collaboration, V.~M. Aulchenko {\em et~al.}, ``{Study of the
  Reactions $e^+ e^- \to \omega \pi^0$ and $e^+ e^- \to \pi^0 \pi^0 \gamma$
  with SND Detector at VEPP-2M},'' BUDKER-INP-2000-35.

\bibitem{CMD-2:2003bgh}
{\bfseries CMD-2} Collaboration, R.~R. Akhmetshin {\em et~al.}, ``{Study of the
  process $e^+ e^- \to \omega \pi^0 \to \pi^0 \pi^0 \gamma$ in c.m. energy
  range 920~MeV -- 1380~MeV at CMD-2},''
  \href{https://dx.doi.org/10.1016/S0370-2693(03)00595-1}{{\em Phys. Lett. B}
  {\bfseries 562} (2003) 173--181},
  \href{https://arxiv.org/abs/hep-ex/0304009}{{\ttfamily
  arXiv:hep-ex/0304009}}.

\bibitem{Achasov:2012zza}
{\bfseries SND} Collaboration, M.~N. Achasov {\em et~al.}, ``{Measurement of
  the cross section for the $e^+ e^- \to \omega \pi^0 \to \pi^0 \pi^0 \gamma$
  process in the energy range of 1.1~GeV -- 1.9~GeV},''
  \href{https://dx.doi.org/10.1134/S0021364011220024}{{\em JETP Lett.}
  {\bfseries 94} (2012) 734--737}.

\bibitem{Achasov:1999wr}
{\bfseries SND} Collaboration, M.~N. Achasov {\em et~al.}, ``{Investigation of
  the $e^+ e^- \to \omega \pi^0 \to \pi^0 \pi^0 \gamma$ reaction in the energy
  domain near the $\phi$ meson},''
  \href{https://dx.doi.org/10.1016/S0550-3213(99)00482-4}{{\em Nucl. Phys. B}
  {\bfseries 569} (2000) 158--182},
  \href{https://arxiv.org/abs/hep-ex/9907026}{{\ttfamily
  arXiv:hep-ex/9907026}}.

\bibitem{Dolinsky:1986kj}
{\bfseries ND} Collaboration, S.~I. Dolinsky {\em et~al.}, ``{The Reaction $e^+
  e^- \to \Omega \pi^0$ in the Center-of-mass Energy Range From 1.0~GeV to
  1.4~GeV},'' \href{https://dx.doi.org/10.1016/0370-2693(86)91036-1}{{\em Phys.
  Lett. B} {\bfseries 174} (1986) 453--457}, NOVOSIBIRSK-86-69.

\bibitem{CMD-2:2003mfj}
{\bfseries CMD-2} Collaboration, R.~R. Akhmetshin {\em et~al.}, ``{Study of the
  process $e^+ e^- \to \pi^0 \pi^0 \gamma$ in c.m. energy range 600~MeV to
  970~MeV at CMD2},''
  \href{https://dx.doi.org/10.1016/j.physletb.2003.11.046}{{\em Phys. Lett. B}
  {\bfseries 580} (2004) 119--128},
  \href{https://arxiv.org/abs/hep-ex/0310012}{{\ttfamily
  arXiv:hep-ex/0310012}}.

\bibitem{Achasov:2016zvn}
{\bfseries SND} Collaboration, M.~N. Achasov {\em et~al.}, ``{Updated
  measurement of the $e^+e^- \to \omega \pi^0 \to \pi^0\pi^0\gamma$ cross
  section with the SND detector},''
  \href{https://dx.doi.org/10.1103/PhysRevD.94.112001}{{\em Phys. Rev. D}
  {\bfseries 94} no.~11, (2016) 112001},
  \href{https://arxiv.org/abs/1610.00235}{{\ttfamily arXiv:1610.00235
  [hep-ex]}}.

\bibitem{Cordier:1981zs}
{\bfseries DM1} Collaboration, A.~Cordier, D.~Bisello, J.~C. Bizot, J.~Buon,
  B.~Delcourt, and F.~Mane, ``{Observation of a New Isoscalar Vector Meson in
  $e^+ e^- \to \omega \pi^+ \pi^-$ Annihilation at 1.65~GeV},''
  \href{https://dx.doi.org/10.1016/0370-2693(81)91101-1}{{\em Phys. Lett. B}
  {\bfseries 106} (1981) 155--158}, LAL 81/20.

\bibitem{Achasov:2016eyg}
{\bfseries SND} Collaboration, M.~N. Achasov {\em et~al.}, ``{Study of the
  process $e^+e^-\to\omega\eta\pi^0$ in the energy range $\sqrt{s} <2$ GeV with
  the SND detector},''
  \href{https://dx.doi.org/10.1103/PhysRevD.94.032010}{{\em Phys. Rev. D}
  {\bfseries 94} no.~3, (2016) 032010},
  \href{https://arxiv.org/abs/1606.06481}{{\ttfamily arXiv:1606.06481
  [hep-ex]}}.

\bibitem{Ivanov:2019crp}
{\bfseries CMD-3} Collaboration, V.~L. Ivanov {\em et~al.}, ``{Study of the
  process $e^+e^-{\to}K^+K^-\eta$ with the CMD-3 detector at the VEPP-2000
  collider},'' \href{https://dx.doi.org/10.1016/j.physletb.2019.134946}{{\em
  Phys. Lett. B} {\bfseries 798} (2019) 134946},
  \href{https://arxiv.org/abs/1906.08006}{{\ttfamily arXiv:1906.08006
  [hep-ex]}}.

\bibitem{BaBar:2018rkc}
{\bfseries BaBar} Collaboration, J.~P. Lees {\em et~al.}, ``{Study of the
  reactions $e^+e^-\to\pi^+\pi^-\pi^0\pi^0\pi^0\gamma$ and
  $\pi^+\pi^-\pi^0\pi^0\eta\gamma$ at center-of-mass energies from threshold to
  4.35 GeV using initial-state radiation},''
  \href{https://dx.doi.org/10.1103/PhysRevD.98.112015}{{\em Phys. Rev. D}
  {\bfseries 98} no.~11, (2018) 112015}, SLAC-PUB-17344,
  \href{https://arxiv.org/abs/1810.11962}{{\ttfamily arXiv:1810.11962
  [hep-ex]}}.

\bibitem{Bisello:1983at}
{\bfseries DM2} Collaboration, D.~Bisello {\em et~al.}, ``{A Measurement of
  $e^+ e^- \to \bar{p} p$ for 1975~{MeV} $\le \sqrt{s} \le$ 2250~{MeV}},''
  \href{https://dx.doi.org/10.1016/0550-3213(83)90381-4}{{\em Nucl. Phys. B}
  {\bfseries 224} (1983) 379}, LAL 83/15.

\bibitem{CMD-3:2015fvi}
{\bfseries CMD-3} Collaboration, R.~R. Akhmetshin {\em et~al.}, ``{Study of the
  process $e^+e^-\to p\bar{p}$ in the c.m. energy range from threshold to 2 GeV
  with the CMD-3 detector},''
  \href{https://dx.doi.org/10.1016/j.physletb.2016.04.048}{{\em Phys. Lett. B}
  {\bfseries 759} (2016) 634--640},
  \href{https://arxiv.org/abs/1507.08013}{{\ttfamily arXiv:1507.08013
  [hep-ex]}}.

\bibitem{Delcourt:1979ed}
{\bfseries DM1} Collaboration, B.~Delcourt {\em et~al.}, ``{Study of the
  Reaction $e^+ e^- \to p \bar{p}$ in the Total Energy Range 1925~MeV --
  2180~MeV},'' \href{https://dx.doi.org/10.1016/0370-2693(79)90864-5}{{\em
  Phys. Lett. B} {\bfseries 86} (1979) 395--398}, LAL-79/12.

\bibitem{BaBar:2013ves}
{\bfseries BaBar} Collaboration, J.~P. Lees {\em et~al.}, ``{Study of $e^+e^-
  \to p \bar{p}$ via initial-state radiation at BABAR},''
  \href{https://dx.doi.org/10.1103/PhysRevD.87.092005}{{\em Phys. Rev. D}
  {\bfseries 87} no.~9, (2013) 092005}, BABAR-PUB-12-030, SLAC-PUB-15324,
  \href{https://arxiv.org/abs/1302.0055}{{\ttfamily arXiv:1302.0055 [hep-ex]}}.

\bibitem{Antonelli:1998fv}
{\bfseries FENICE} Collaboration, A.~Antonelli {\em et~al.}, ``{The first
  measurement of the neutron electromagnetic form-factors in the timelike
  region},'' \href{https://dx.doi.org/10.1016/S0550-3213(98)00083-2}{{\em Nucl.
  Phys. B} {\bfseries 517} (1998) 3--35}.

\bibitem{Achasov:2014ncd}
{\bfseries SND} Collaboration, M.~N. Achasov {\em et~al.}, ``{Study of the
  process $e^+e^-\to n\bar{n}$ at the VEPP-2000 $e^+e^-$ collider with the SND
  detector},'' \href{https://dx.doi.org/10.1103/PhysRevD.90.112007}{{\em Phys.
  Rev. D} {\bfseries 90} no.~11, (2014) 112007},
  \href{https://arxiv.org/abs/1410.3188}{{\ttfamily arXiv:1410.3188 [hep-ex]}}.

\bibitem{BES:2009ejh}
{\bfseries BES} Collaboration, M.~Ablikim {\em et~al.}, ``{R value measurements
  for $e^+ e^-$ annihilation at 2.60~GeV, 3.07~GeV and 3.65~GeV},''
  \href{https://dx.doi.org/10.1016/j.physletb.2009.05.055}{{\em Phys. Lett. B}
  {\bfseries 677} (2009) 239--245},
  \href{https://arxiv.org/abs/0903.0900}{{\ttfamily arXiv:0903.0900 [hep-ex]}}.

\bibitem{BES:2001ckj}
{\bfseries BES} Collaboration, J.~Z. Bai {\em et~al.}, ``{Measurements of the
  cross-section for $e^+ e^- \to hadrons$ at center-of-mass energies from 2~GeV
  to 5~GeV},'' \href{https://dx.doi.org/10.1103/PhysRevLett.88.101802}{{\em
  Phys. Rev. Lett.} {\bfseries 88} (2002) 101802}, SLAC-PUB-8938,
  \href{https://arxiv.org/abs/hep-ex/0102003}{{\ttfamily
  arXiv:hep-ex/0102003}}.

\bibitem{Ablikim:2006mb}
{\bfseries BES II} Collaboration, M.~Ablikim {\em et~al.}, ``{Measurements of
  the continuum R(uds) and R values in $e^+ e^-$ annihilation in the energy
  region between 3.650 and 3.872~GeV},''
  \href{https://dx.doi.org/10.1103/PhysRevLett.97.262001}{{\em Phys. Rev.
  Lett.} {\bfseries 97} (2006) 262001},
  \href{https://arxiv.org/abs/hep-ex/0612054}{{\ttfamily
  arXiv:hep-ex/0612054}}.

\bibitem{Barber:1981tc}
{\bfseries Mark-J} Collaboration, D.~P. Barber {\em et~al.}, ``{Measurement of
  Hadron Production and Three Jet Event Properties at {PETRA} {MIT}},''
  \href{https://dx.doi.org/10.1016/0370-2693(82)91143-1}{{\em Phys. Lett. B}
  {\bfseries 108} (1982) 63--66}, MIT-LNS-115.

\bibitem{Albrecht:1982bq}
{\bfseries DASP} Collaboration, H.~Albrecht {\em et~al.}, ``{The Hadronic
  Cross-section of Electron - Positron Annihilation at 9.5~GeV and the
  $\Upsilon$ and $\Upsilon^\prime$ Resonance Parameters},''
  \href{https://dx.doi.org/10.1016/0370-2693(82)90305-7}{{\em Phys. Lett. B}
  {\bfseries 116} (1982) 383--386}, DESY-82-037.

\bibitem{TASSO:1984dcd}
{\bfseries TASSO} Collaboration, M.~Althoff {\em et~al.}, ``{Measurement of r
  and Search for the Top Quark in $e^+ e^-$ Annihilation Between 39.8~GeV and
  45.2~GeV},'' \href{https://dx.doi.org/10.1016/0370-2693(84)91936-1}{{\em
  Phys. Lett. B} {\bfseries 138} (1984) 441--448}, DESY-84-001.

\bibitem{CELLO:1986ijz}
{\bfseries CELLO} Collaboration, H.~J. Behrend {\em et~al.}, ``{Determination
  of alpha-s and sin**2theta(w) from Measurements of the Total Hadronic
  Cross-Section in $e^+ e^-$ Annihilation},''
  \href{https://dx.doi.org/10.1016/0370-2693(87)90986-5}{{\em Phys. Lett. B}
  {\bfseries 183} (1987) 400--411}, DESY-86-133.

\bibitem{Venus:1987dew}
{\bfseries VENUS} Collaboration, H.~Yoshida {\em et~al.}, ``{Measurement of R
  and Search for New Heavy Quarks in $e^+ e^-$ Annihilation at 50~GeV and
  52~GeV Center-of-mass Energies},''
  \href{https://dx.doi.org/10.1016/0370-2693(87)90920-8}{{\em Phys. Lett. B}
  {\bfseries 198} (1987) 570}, KEK-PREPRINT-87-81, KOBE-HEP-87-03, KUNS-893,
  OULNS-87-5, TMUP-HEL-87-22.

\bibitem{TOPAZ:1989phk}
{\bfseries TOPAZ} Collaboration, I.~Adachi {\em et~al.}, ``{Measurement of the
  Total Hadronic Cross-sections in e$^{+} $e$^{-}$ Annihilation and
  Determination of the Standard Model Parameters},''
  \href{https://dx.doi.org/10.1016/0370-2693(90)92052-K}{{\em Phys. Lett. B}
  {\bfseries 234} (1990) 525--533}, KEK-PREPRINT-89-136, DPNU-89-51, INS-779,
  KOBE-HEP-89-10, NWU-HEP-89-04, OCU-HEP-89-04, PU-89-636, TU-HEP-89-03,
  TUAT-HEP-89-03, UT-HE-89-08.

\bibitem{VENUS:1990vwh}
{\bfseries VENUS} Collaboration, K.~Abe {\em et~al.}, ``{Experimental limits on
  extra $Z$ bosons from $e^{+} e^{-}$ annihilation data with the VENUS detector
  at $\sqrt{s}$ = 50~GeV to approximately 64~GeV},''
  \href{https://dx.doi.org/10.1016/0370-2693(90)91348-F}{{\em Phys. Lett. B}
  {\bfseries 246} (1990) 297--305}, KEK-PREPRINT-90-34.

\bibitem{TOPAZ:1993iqj}
{\bfseries TOPAZ} Collaboration, K.~Abe {\em et~al.}, ``{Search for a narrow
  resonance in $e^+ e^-$ collisions between E(cm) = 58~GeV and 60~GeV},''
  \href{https://dx.doi.org/10.1016/0370-2693(93)90311-5}{{\em Phys. Lett. B}
  {\bfseries 304} (1993) 373--380}, KEK-PREPRINT-92-205, TUAT-HEP-93-01,
  NWU-HEP-9302, DPNU-93-08, OCU-HEP-93-1, PU-93-666, INS-966, KOBE-HEP-93-04,
  TIT-HPE-93-03.

\bibitem{TOPAZ:1994ymb}
{\bfseries TOPAZ} Collaboration, K.~Miyabayashi {\em et~al.}, ``{Measurement of
  the total hadronic cross-section and determination of gamma - Z interference
  in $e^+ e^-$ annihilation},''
  \href{https://dx.doi.org/10.1016/0370-2693(95)00038-M}{{\em Phys. Lett. B}
  {\bfseries 347} (1995) 171--178}, DPNU-94-47, NWU-HEP-94-06,
  KEK-PREPRINT-94-152, TUAT-HEP-94-05, TIT-HPE-94-11, OCU-HEP-94-09, PU-94-688,
  INS-1074, KOBE-HEP-94-08.

\bibitem{KEDR:2018hhr}
{\bfseries KEDR} Collaboration, V.~V. Anashin {\em et~al.}, ``{Precise
  measurement of $R_{\text{uds}}$ and $R$ between 1.84 and 3.72 GeV at the KEDR
  detector},'' \href{https://dx.doi.org/10.1016/j.physletb.2018.11.012}{{\em
  Phys. Lett. B} {\bfseries 788} (2019) 42--51},
  \href{https://arxiv.org/abs/1805.06235}{{\ttfamily arXiv:1805.06235
  [hep-ex]}}.

\bibitem{Aachen-DESY-AnnecyLAPP-MIT-NIKHEF-Beijing:1979sxf}
{\bfseries Aachen-DESY-Annecy(LAPP)-MIT-NIKHEF-Beijing} Collaboration, D.~P.
  Barber {\em et~al.}, ``{Study of Electron - Positron Collisions at the
  Highest {PETRA} Energy},''
  \href{https://dx.doi.org/10.1016/0370-2693(79)91296-6}{{\em Phys. Lett. B}
  {\bfseries 85} (1979) 463--466}, MIT-LNS-104.

\bibitem{CLEO:1983ffg}
{\bfseries CLEO} Collaboration, R.~Giles {\em et~al.}, ``{The Total
  Cross-section for Electron - Positron Annihilation Into Hadron Final States
  in the $\Upsilon$ Energy Region},''
  \href{https://dx.doi.org/10.1103/PhysRevD.29.1285}{{\em Phys. Rev. D}
  {\bfseries 29} (1984) 1285}, CLNS-83/586, CLEO-83-10.

\bibitem{Fernandez:1984yw}
{\bfseries SLAC-PEP-MAC} Collaboration, E.~Fernandez {\em et~al.}, ``{Precision
  Measurement of the Total Cross-section for $e^+ e^- \to$ Hadrons at a
  Center-of-mass Energy of 29~GeV},''
  \href{https://dx.doi.org/10.1103/PhysRevD.31.1537}{{\em Phys. Rev. D}
  {\bfseries 31} (1985) 1537}, SLAC-PUB-3479.

\bibitem{Mark-J:1986ave}
{\bfseries Mark-J} Collaboration, B.~Adeva {\em et~al.}, ``{Study of Hadron and
  Inclusive Muon Production From Electron Positron Annihilation at 39.79~GeV
  $\le \sqrt{s} \le$ 46.78~GeV},''
  \href{https://dx.doi.org/10.1103/PhysRevD.34.681}{{\em Phys. Rev. D}
  {\bfseries 34} (1986) 681--691}, MIT-LNS-146.

\bibitem{AMY:1990xpt}
{\bfseries AMY} Collaboration, T.~Kumita {\em et~al.}, ``{Measurements of $R$
  for $e^+e^-$ annihilation at TRISTAN},''
  \href{https://dx.doi.org/10.1103/PhysRevD.42.1339}{{\em Phys. Rev. D}
  {\bfseries 42} (1990) 1339--1349}, KEK-PREPRINT-89-188, AMY-89-21.

\bibitem{MARK-II:1990awn}
{\bfseries MARK-II} Collaboration, C.~Von~Zanthier {\em et~al.}, ``{Measurement
  of the Total Hadronic Cross-section in $e^+ e^-$ Annihilation at
  $\sqrt{s}=29$~GeV},'' \href{https://dx.doi.org/10.1103/PhysRevD.43.34}{{\em
  Phys. Rev. D} {\bfseries 43} (1991) 34--45}, SLAC-PUB-5213, LBL-28859.

\bibitem{CLEO:1997eca}
{\bfseries CLEO} Collaboration, R.~Ammar {\em et~al.}, ``{A Measurement of the
  total cross-section for $e^+ e^- \to hadrons$ at $\sqrt{s} = 10.52$~GeV},''
  \href{https://dx.doi.org/10.1103/PhysRevD.57.1350}{{\em Phys. Rev. D}
  {\bfseries 57} (1998) 1350--1358}, SLAC-PUB-9765, CLNS-97-1493, CLEO-97-14,
  \href{https://arxiv.org/abs/hep-ex/9707018}{{\ttfamily
  arXiv:hep-ex/9707018}}.

\bibitem{Rapidis:1977cv}
{\bfseries SLAC-SPEAR} Collaboration, P.~A. Rapidis {\em et~al.},
  ``{Observation of a Resonance in $e^+ e^-$ Annihilation Just Above Charm
  Threshold},'' \href{https://dx.doi.org/10.1103/PhysRevLett.39.526}{{\em Phys.
  Rev. Lett.} {\bfseries 39} (1977) 526}, SLAC-PUB-1959, LBL-6484. [Erratum:
  Phys.Rev.Lett. 39, 974 (1977)].

\bibitem{Rice:1982br}
{\bfseries CUSB} Collaboration, E.~Rice {\em et~al.}, ``{Search for structure
  in $\sigma(e^+ e^- \to hadrons)$ between $\sqrt{s} = 10.34$~GeV and
  $11.6$~GeV},'' \href{https://dx.doi.org/10.1103/PhysRevLett.48.906}{{\em
  Phys. Rev. Lett.} {\bfseries 48} (1982) 906--910}.

\bibitem{BES:1999wbx}
{\bfseries BES} Collaboration, J.~Z. Bai {\em et~al.}, ``{Measurement of the
  total cross-section for hadronic production by $e^+ e^-$ annihilation at
  energies between 2.6~GeV -- 5~GeV},''
  \href{https://dx.doi.org/10.1103/PhysRevLett.84.594}{{\em Phys. Rev. Lett.}
  {\bfseries 84} (2000) 594--597}, SLAC-REPRINT-1999-087,
  \href{https://arxiv.org/abs/hep-ex/9908046}{{\ttfamily
  arXiv:hep-ex/9908046}}.

\bibitem{Naroska:1986si}
{\bfseries JADE} Collaboration, B.~Naroska, ``{$e^+ e^-$ Physics with the JADE
  Detector at PETRA},''
  \href{https://dx.doi.org/10.1016/0370-1573(87)90031-7}{{\em Phys. Rept.}
  {\bfseries 148} (1987) 67}, DESY-86-113.

\bibitem{Schindler:1979rj}
{\bfseries Mark-II} Collaboration, R.~H. Schindler, ``{Charmed meson production
  and decay properties at the $\Psi''(3770)$},'' SLAC-R-0219.

\bibitem{Osterheld:1986hw}
{\bfseries SLAC-SPEAR} Collaboration, A.~Osterheld {\em et~al.},
  ``{Measurements of Total Hadronic and Inclusive $D^*$ Cross-sections in $e^+
  e^-$ Annihilations Between 3.87~GeV and 4.5~GeV},'' SLAC-PUB-4160.

\bibitem{Edwards:1990pc}
{\bfseries Crystal Ball} Collaboration, C.~Edwards {\em et~al.}, ``{Hadron
  Production in $e^+ e^-$ Annihilation From $\sqrt{s}=5$~GeV to 7.4~GeV},''
  SLAC-PUB-5160.

\bibitem{LENA:1982shw}
{\bfseries LENA} Collaboration, B.~Niczyporuk {\em et~al.}, ``{Measurement of R
  in $e^+ e^-$ Annihilation for $\sqrt{s}$ Between 7.4~GeV and 9.4~GeV},''
  \href{https://dx.doi.org/10.1007/BF01614421}{{\em Z. Phys. C} {\bfseries 15}
  (1982) 299}, DESY-82-052.

\bibitem{TASSO:1983cre}
{\bfseries TASSO} Collaboration, M.~Althoff {\em et~al.}, ``{Jet Production and
  Fragmentation in $e^+ e^-$ Annihilation at 12~GeV to 43~GeV},''
  \href{https://dx.doi.org/10.1007/BF01547419}{{\em Z. Phys. C} {\bfseries 22}
  (1984) 307--340}, DESY-83-130.

\bibitem{CrystalBall:1988obw}
{\bfseries Crystal Ball} Collaboration, Z.~Jakubowski {\em et~al.},
  ``{Determination of $\Gamma(ee$) of the $\Upsilon$ (1s) and $\Upsilon$ (2s)
  Resonances and Measurement of R at $W$=9.39~GeV},''
  \href{https://dx.doi.org/10.1007/BF01559717}{{\em Z. Phys. C} {\bfseries 40}
  (1988) 49}, SLAC-PUB-4567, DESY-88-032.

\bibitem{TASSO:1990cdg}
{\bfseries TASSO} Collaboration, W.~Braunschweig {\em et~al.}, ``{Global Jet
  Properties at 14~GeV to 44~GeV Center-of-mass Energy in $e^+ e^-$
  Annihilation},'' \href{https://dx.doi.org/10.1007/BF01552339}{{\em Z. Phys.
  C} {\bfseries 47} (1990) 187--198}, DESY-90-013.

\bibitem{TASSO:1979rma}
{\bfseries TASSO} Collaboration, R.~Brandelik {\em et~al.}, ``{$e^+ e^-$
  Annihilation at High-Energies and Search for the t-Quark Continuum
  Contribution},'' \href{https://dx.doi.org/10.1007/BF01554391}{{\em Z. Phys.
  C} {\bfseries 4} (1980) 87}, DESY-79/74.

\bibitem{ARGUS:1991ztk}
{\bfseries ARGUS} Collaboration, H.~Albrecht {\em et~al.}, ``{Measurement of R
  and determination of the charged particle multiplicity in $e^+ e^-$
  annihilation at $\sqrt{s}$ around 10~GeV},''
  \href{https://dx.doi.org/10.1007/BF01881704}{{\em Z. Phys. C} {\bfseries 54}
  (1992) 13--20}, DESY-91-092.

\bibitem{DESY-Hamburg-Heidelberg-Munich:1980ttv}
{\bfseries DESY-Hamburg-Heidelberg-Munich} Collaboration, P.~Bock {\em et~al.},
  ``{Total Cross-section for Hadron Production by $e^+ e^-$ Annihilation
  Between 9.4~GeV and 9.5~GeV},''
  \href{https://dx.doi.org/10.1007/BF01588838}{{\em Z. Phys. C} {\bfseries 6}
  (1980) 125}, DESY-80-58.

\bibitem{Blinov:1993fw}
{\bfseries MD-1} Collaboration, A.~E. Blinov {\em et~al.}, ``{The Measurement
  of $R$ in $e^+ e^-$ annihilation at center-of-mass energies between 7.2~GeV
  and 10.34~GeV},'' \href{https://dx.doi.org/10.1007/s002880050077}{{\em Z.
  Phys. C} {\bfseries 70} (1996) 31--38}, BUDKER-INP-1993-54.

\bibitem{Borsanyi:2020mff}
S.~Borsanyi {\em et~al.}, ``{Leading hadronic contribution to the muon magnetic
  moment from lattice QCD},''
  \href{https://dx.doi.org/10.1038/s41586-021-03418-1}{{\em Nature} {\bfseries
  593} no.~7857, (2021) 51--55},
  \href{https://arxiv.org/abs/2002.12347}{{\ttfamily arXiv:2002.12347
  [hep-lat]}}.

\end{thebibliography}\endgroup

\end{document}